\documentclass{article} % For LaTeX2e
\usepackage{arxiv,times}

\usepackage[super,sort&compress,comma]{natbib} 
\usepackage{balance}
\usepackage{mathptmx}
\usepackage{sectsty}
\usepackage{graphicx}
\usepackage{adjustbox}
\usepackage{tabularx}
\usepackage{multirow}
\usepackage{rotating}
\usepackage{booktabs}
\usepackage{placeins}
\usepackage{amsmath,amssymb}
\usepackage{bm}
\usepackage{graphicx}
\usepackage{adjustbox}
\usepackage{tabularx}
\usepackage{multirow}
\usepackage{rotating}
\usepackage{booktabs}
\usepackage{placeins}
\newcommand{\C}{\mathcal{C}}
\newcommand{\N}{\mathcal{N}}

\newcommand{\angstrom}{\text{\normalfont\AA}}

\newcommand{\ic}{IC$_{\text{50}}$}
\newcommand{\pilot}{PILOT}

\newcommand{\denovo}{\textit{de novo}}
\usepackage{lastpage}
\usepackage[format=plain,justification=justified,singlelinecheck=false,font={stretch=1.125,small,sf},labelfont=bf,labelsep=space]{caption}
\usepackage{float}
\usepackage{fancyhdr}
\usepackage{fnpos}
\usepackage[english]{babel}
\addto{\captionsenglish}{%
  \renewcommand{\refname}{References}
}
\usepackage{array}
\usepackage{droidsans}
\usepackage{charter}
\usepackage[T1]{fontenc}
\usepackage[usenames,dvipsnames]{xcolor}
\usepackage{setspace}
\usepackage[compact]{titlesec}
\usepackage{hyperref}
\usepackage{algorithm}% http://ctan.org/pkg/algorithms
\usepackage{algpseudocode}% http://ctan.org/pkg/algorithmicx

\usepackage{epstopdf}%This line makes .eps figures into .pdf - please comment out if not required.

\definecolor{cream}{RGB}{222,217,201}

\title{PILOT: Equivariant diffusion for pocket conditioned de novo ligand generation with multi-objective guidance via importance sampling}

\author{Julian Cremer$^*$ \\
Pfizer Research \& Development\\
University Pompeu Fabra\\
\texttt{julian.cremer@pfizer.com} \\
\And
Tuan Le\thanks{Shared co-first authorship} \\
Pfizer Research \& Development\\
Freie Universität Berlin \\
\texttt{tuan.le@pfizer.com} \\
\And
Frank Noé \\
Freie Universität Berlin \\
Microsoft Research\\
\texttt{frank.noe@microsoft.com} \\
\And
Djork-Arné Clevert \\
Pfizer Research \& Development\\
\texttt{djork-arne.clevert@pfizer.com} \\
\And
Kristof T. Schütt \\
Pfizer Research \& Development\\
\texttt{kristof.schuett@pfizer.com} \\
}

\begin{document}

\maketitle

\begin{abstract}
The generation of ligands that both are tailored to a given protein pocket and exhibit a range of desired chemical properties is a major challenge in structure-based drug design.
Here, we propose an in-silico approach for the \textit{de novo} generation of 3D ligand structures using the equivariant diffusion model \pilot, combining pocket conditioning with a large-scale pre-training and property guidance.
Its multi-objective trajectory-based importance sampling strategy is designed to direct the model towards molecules that not only exhibit desired characteristics such as increased binding affinity for a given protein pocket but also maintains high synthetic accessibility.
This ensures the practicality of sampled molecules, thus maximizing their potential for the drug discovery pipeline.
\pilot~significantly outperforms existing methods across various metrics on the common benchmark dataset CrossDocked2020. 
Moreover, we employ \pilot~to generate novel ligands for unseen protein pockets from the Kinodata-3D dataset, which encompasses a substantial portion of the human kinome.
The generated structures exhibit predicted \ic~values indicative of potent biological activity, which highlights the potential of \pilot~as a powerful tool for structure-based drug design.
\end{abstract}

\section{Introduction}
Structure-based drug discovery (SBDD) has fundamentally transformed the landscape of drug development by facilitating the design of molecules that precisely target biological macromolecules, such as proteins, which play a critical role in disease processes. These designed molecules interact with a specific pocket of a target protein, either activating or inhibiting its function, thus influencing the disease pathway. This strategy is underpinned by a detailed understanding of the 3D structure of the target, usually acquired through X-ray crystallography or nuclear magnetic resonance (NMR) spectroscopy. \cite{Anderson2003, Batool2019} 
By grasping the structural intricacies of the target protein, researchers are equipped to create ligands that specifically  modulate its activity, offering potential therapeutic benefits.

A major challenge in SBDD is the vast chemical space that must be navigated to discover molecules with desired properties. 
%For example, high-throughput screening requires substantial time and cost and may still yield numerous compounds that are unsuitable for further development.
Recently, machine learning (ML) has been applied to SBDD, promising to enable researchers to rapidly pinpoint drug candidates, significantly reducing the reliance on labor-intensive and costly experimental methods. 
ML algorithms are capable of analyzing vast datasets of molecular structures and properties to discern patterns, predict outcomes and generate \denovo~molecules.
This might not only accelerate the drug discovery process but also enhance the efficiency and efficacy of identifying viable therapeutic agents. \cite{deepfrag_2021, luo_2021, Ragoza2022, liu2022, tan2022, peng_2022, dror_2023}

One of the innovative machine learning techniques increasingly employed in structure-based drug discovery is the application of generative diffusion models. 
Originally utilized in fields like computer vision and natural language processing, these models also excel in capturing the complex patterns of 3D molecular structures, particularly when enhanced with features that reflect the symmetry and specific target-related characteristics of proteins. \cite{luo_2021, peng_2022, guan2023d, schneuing_2023_structurebased} 
Another line of research leverages diffusion models as methodology to build ML based docking models. \cite{corso_2023, luhua_2024}

The effectiveness of these models hinges on training with detailed protein structures, which allows for the generation of ligands that are not only structurally compatible but also specifically designed for the interaction with target proteins. However, while generated ligands fit well in a protein binding pocket, these methods lack a mechanism to guide the generative process towards ligands with desired chemical properties such as binding affinity, stability, or bioavailability. Additionally, 3D generative models often yield ligands with a high prevalence of fused rings and low synthetic accessibility.\cite{tamgen, schneuing_2023_structurebased, guan2023d, bombarelli_vae,  winter_mso_2019}

\begin{figure*}[htb!]
  \centering
  \includegraphics[width=0.8\textwidth]{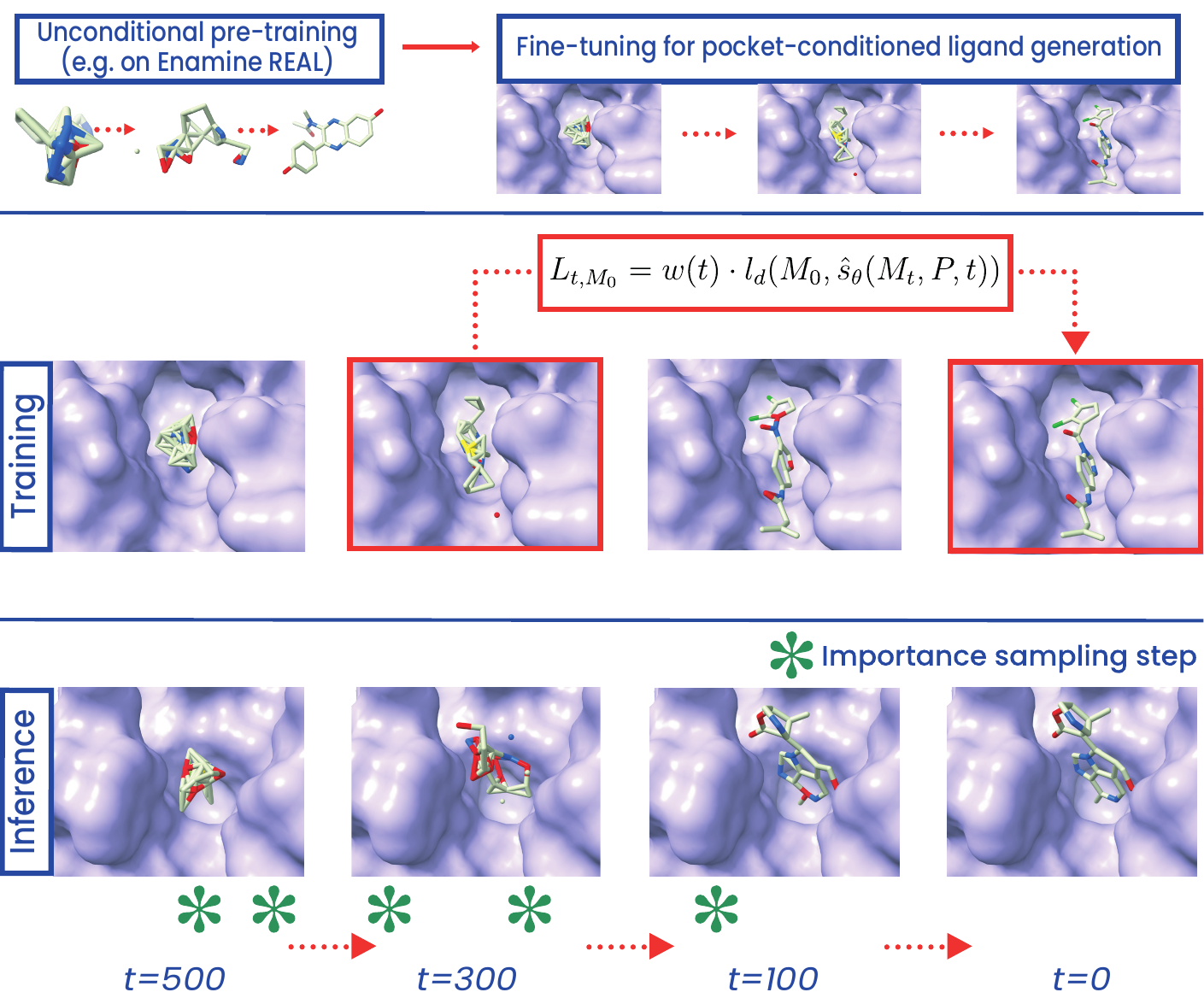}
  \caption{
  \textbf{Top}: \pilot $ $  is first pre-trained unconditionally on an Enamine Real subset from the ZINC database. \citep{irwin_zinc15} We employ OpenEye's Omega to create at most five conformers per molecule.\citep{hawkins_2011_oeomega} Afterwards, we fine-tune the model on CrossDocked2020 conditioned on the atoms of the pocket.\citep{crossdocked}. \textbf{Middle}: Given the binding pocket of a protein, a noisy state of a ligand is sampled from the diffusion forward trajectory (here, t=300) as input to the diffusion model during training. The model has to retrieve the ground truth ligand ($M_0$). For training, a composite loss ($l_d$) is used for continuous (mean squared error) and categorical features (cross-entropy loss), respectively, together with a timestep-dependent loss weighting ($w(t)$). \textbf{Bottom}: At inference, a point cloud is sampled from a Gaussian prior (t=500). Given a binding pocket, the model retrieves a fitting ligand by following the reverse diffusion trajectory. At pre-specified steps, a property surrogate model (green crosses) guides the diffusion process towards desired regions in chemical space using importance sampling.}
  \label{fig:graphical-abstract-method}
\end{figure*}

In this study, we introduce \pilot $ $ (\textbf{P}ocket-induced \textbf{L}igand \textbf{O}ptimization \textbf{T}ool) $ $ -- an equivariant diffusion model designed for de novo ligand generation. As shown in Fig.\ref{fig:graphical-abstract-method}, \pilot $ $ operates in three distinct stages: unconditional diffusion pre-training, pocket conditioned fine-tuning, and property-guided inference. During the inference stage, we employ an importance sampling scheme to replace less desirable intermediate samples with more favorable ones, thereby re-weighting trajectories during generation.
This strategy enables the use of any pre-trained, unconditioned diffusion score model for sampling, which is subsequently enhanced by integrating the capabilities of an external model, similar to classifier guidance.\citep{dhariwal2021diffusion}
However, while classifier guidance may drive the sampling trajectory to adversarial, out-of-distribution structures, \citep{dhariwal2021diffusion} trajectory re-weighting ensures that samples remain within distribution. 
%To prevent mode collapse, importance sampling is applied selectively at every n-th step instead of the entire trajectory. 
As trajectory re-weighting can be conducted in parallel for multiple properties, we focus on three critical properties for drug discovery: synthetic accessibility (SA), docking score, and potency (IC$_{\text{50}}$). 
Our findings demonstrate that \pilot $ $ generates ligands that not only exhibit a significant improvement in synthesizability and drug-likeness but also achieve favorable docking scores and predicted inhibition.

%\section{Method}
\section{Results and discussion}

\subsection{Pre-training of pocket conditioned 3D diffusion models}
\begin{figure*}[htb!]
\centering
  \includegraphics[width=0.8\textwidth]{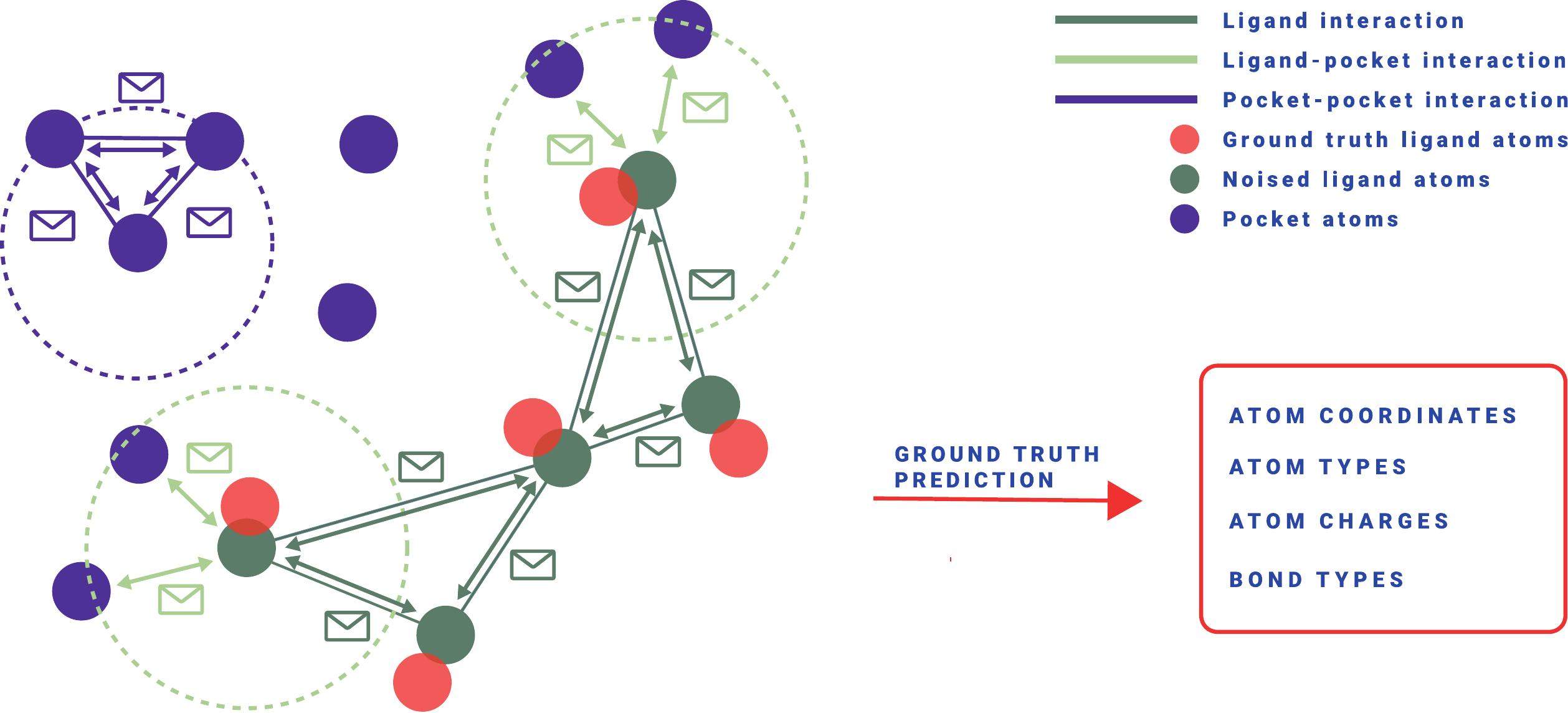}
  \caption{Schematical depiction of the \pilot $ $  network. Given fixed pocket atoms (purple), ligand atom coordinates, types, and charges as well as the ligands' topology get noised (green) using forward diffusion. Afterwards, attention-weighted message-passing is done on the fully connected ligand atoms (here not shown for better visibility) and the ligand-pocket and pocket-pocket interactions, which each are obtained using a radius graph for computational feasibility. The task of the model is to retrieve the ground truth atom coordinates, types, charges, and the bond types (red).}
  \label{fig:eqgat_diff_pocket}
\end{figure*}

%\begin{figure*}[htb]
%\centering
%   \includegraphics[width=0.8\textwidth]{figures/Grapf abstract4.pdf}
%   \caption{\pilot $ $  is first pre-trained unconditionally on an Enamine Real subset from the ZINC database. \citep{irwin_zinc15} We employ OpenEye's Omega to create at most five conformers per molecule.\citep{hawkins_2011_oeomega} Afterwards, we fine-tune the model on CrossDocked2020 conditioned on the atoms of the pocket.\citep{crossdocked}}
%   \label{fig:finetune}
% \end{figure*}

Pre-training enables deep neural networks to build efficient representations by learning the underlying structure of the data.
It proves to be a successful strategy across various fields of machine learning, particularly in the development of large language models (LLMs).\citep{gpt3, bert}
The success of these methods provides a compelling case for applying similar methodologies in the domain of scientific research, specifically in computational chemistry and drug discovery.\citep{winter_cddd_2019, liu2023molecular, zaidi2023pretraining} 
% In the realm of natural language processing (NLP), the concept of pre-training involves training a model on a vast and diverse corpus of text before fine-tuning it on a more specific task or dataset.
% Pre-training allows these models to learn a rich representation of language, capturing nuances, syntax, and semantics that are universally applicable across various tasks.
In the context of \denovo~molecular diffusion models, pre-training allows the models to learn fundamental chemical properties and interactions from large datasets of molecular structures.
This foundational knowledge includes understanding bond types, molecular geometries, and basic physicochemical properties, which are critical for predicting how novel molecules might interact with biological targets.

Pre-training molecular diffusion models on extensive datasets of low-fidelity 3D molecule data is a beneficial strategy for enhancing \denovo~molecule generation capabilities. 
It significantly enhances the ability of the model to generate structurally diverse and chemically plausible molecules, when subsequently fine-tuned on smaller, high-fidelity datasets.\citep{le_cremer2023}
In this work, we train \pilot $ $  as illustrated in Fig. \ref{fig:eqgat_diff_pocket}. 
For pre-training, we utilize the Enamine Real Diversity subset present in the ZINC database \citep{irwin_zinc15} which we downloaded from the Enamine website. To prepare the dataset, we employ OpenEye's Omega software \citep{hawkins_2011_oeomega}, which we use for the creation of up to five conformers per molecule, resulting in a substantial corpus of approximately 109 million 3D structures. Additionally, we simplify the molecular representations by removing explicit hydrogens.

We study the impact of pre-training on molecules and fine-tuning on ligand-pocket complexes on model performance using the CrossDocked2020 dataset~\citep{luo_2021} following the methodologies described in \citet{le_cremer2023} and detailed in Section \ref{sec:methods}. 
Table \ref{tab:finetuning} shows the results with evaluation of metrics related to the generated molecular structures, such as molecular validity, the number of connected components, and the distribution of bond angles and lengths.
This includes a comparison between models trained from scratch and those that have been pre-trained. 
The chosen distance cutoff of the pocket-ligand complex is a critical factor for model performance in terms of computational cost and accuracy (see Section~\ref{sec:cutoff}).
We find that pre-training improves our models across all measured metrics, and the pre-trained model with 7 {\AA} cutoff achieves state-of-the-art performance for 8 out of the 9 evaluated metrics.
In particular, over 98\% of molecules sampled by the model are PoseBusters-valid (compared to 81\% by TargetDiff). We measure PoseBusters-validity by summing over all non-overlapping evaluations of the "dock" and "mol" configuration in the PoseBusters tool and divide by the number of evaluations.\cite{buttenschoen2023}
The model achieves a Wasserstein distance error of 2.39$_{\pm{0.98}}$ for bond angles. 
This constitutes ~4x improvement over TargetDiff, which indicates a markedly improved ability to learn the underlying data distribution.
Beyond that, all \pilot $ $  models outperform TargetDiff in quantitative estimates of drug-likeness (QED) and synthetic accessibility (SA) scores, indicating that \pilot $ $  not only generates more structurally accurate molecules but also produces compounds that are more drug-like and better synthesizable.

\begin{table*}[htb!]
\caption{Diverse set of evaluation metrics on the CrossDocked2020 test set comprising 100 protein pockets to assess the distribution learning capability. For each protein pocket, 100 ligands are sampled. We compare metrics including novelty, BondLengthsW$_1$, and BondAnglesW$_1$ with respect to the test set. The results are reported as mean values across all targets and ligands, with the standard deviation noted in the subscript.}
\begin{adjustbox}{width=1.0\textwidth,center}
\begin{tabular}{@{}lrrrrrrr}
\toprule
\textbf{Model} &  \pilot$_\text{pocket, 5A}^{\text{scratch}}$ & \pilot$_\text{pocket, 5A}^{\text{pre-train}}$ & \pilot$_\text{pocket, 6A}^{\text{scratch}}$ & \pilot$_\text{pocket, 6A}^{\text{pre-train}}$ & \pilot$_\text{pocket, 7A}^{\text{scratch}}$ & \pilot$_\text{pocket, 7A}^{\text{pre-train}}$ & TargetDiff$_\text{10A}$\\
\midrule
Validity $\uparrow$ & 93.40$_{\pm{5.11}}$ & \textbf{96.08}$_{\pm{3.53}}$ & 93.48$_{\pm{5.13}}$ & 95.47$_{\pm{3.91}}$ & 92.06$_{\pm{6.26}}$  & \textbf{96.05}$_{\pm{3.83}}$ & 78.91$_{\pm{2.45}}$ \\
PoseBusters-valid $\uparrow$ & 96.93$_{\pm{1.91}}$ & 97.39$_{\pm{1.58}}$ & 97.88$_{\pm{1.41}}$ & 97.49$_{\pm{1.72}}$ & 96.92$_{\pm{1.91}}$ & \textbf{98.21}$_{\pm{1.51}}$ & 80.53$_{\pm{1.21}}$ \\
Connect. Comp. $\uparrow$ &  95.61$_{\pm{4.15}}$ & 97.44$_{\pm{2.66}}$ & 95.04$_{\pm{5.02}}$ & 97.19$_{\pm{3.38}}$ & 93.96$_{\pm{5.99}}$ & \textbf{97.81}$_{\pm{3.18}}$ & 88.02$_{\pm{2.54}}$ \\
Diversity $\uparrow$ & 72.12$_{\pm{9.05}}$ & 72.99$_{\pm{9.01}}$ & 72.03$_{\pm{9.48}}$  & 71.66$_{\pm{9.79}}$  & 70.36$_{\pm{9.59}}$ &  71.52$_{\pm{9.84}}$ & \textbf{75.12}$_{\pm{6.41}}$ \\
QED $\uparrow$&  0.50$_{\pm{0.12}}$ & 0.51 $_{\pm{0.12}}$ & 0.51$_{\pm{0.14}}$  & \textbf{0.53}$_{\pm{0.13}}$ & 0.49$_{\pm{0.14}}$ &  \textbf{0.53}$_{\pm{0.12}}$ & 0.42$_{\pm{0.09}}$ \\
SA $\uparrow$&  0.67$_{\pm{0.08}}$ & \textbf{0.69}$_{\pm{0.07}}$ &0.66$_{\pm{0.09}}$  & \textbf{0.69}$_{\pm{0.07}}$ & 0.66$_{\pm{0.07}}$  &  \textbf{0.69}$_{\pm{0.06}}$ & 0.61$_{\pm{0.06}}$ \\
Lipinski $\uparrow$&  4.53$_{\pm{0.53}}$ & 4.54$_{\pm{0.49}}$ &4.54$_{\pm{0.61}}$  & 4.57$_{\pm{0.56}}$  &  4.46$_{\pm{0.65}}$ &  \textbf{4.60}$_{\pm{0.51}}$ & \textbf{4.64}$_{\pm{0.31}}$\\
BondAnglesW$_1$ $\downarrow$ &  4.03$_{\pm{1.29}}$ & 3.04$_{\pm{1.19}}$ & 3.47$_{\pm{1.02}}$  &  3.09$_{\pm{1.06}}$  & 4.00$_{\pm{1.10}}$ & \textbf{2.39}$_{\pm{0.98}}$ & 9.71$_{\pm{4.67}}$ \\
BondLenghtsW$_1$ [$10^{-2}$] $\downarrow$ &  0.27$_{\pm{0.01}}$ & 0.24$_{\pm{0.007}}$ & 0.27$_{\pm{0.09}}$  &  0.23$_{\pm{0.08}}$ &  0.29$_{\pm{0.09}}$ &  \textbf{0.21}$_{\pm{0.08}}$ & 5.12$_{\pm{2.05}}$ \\
Ligand size &  23.70$_{\pm{8.80}}$ & 24.08$_{\pm{8.83}}$ & 24.56$_{\pm{8.81}}$  &  24.70$_{\pm{8.74}}$ &  24.39$_{\pm{8.74}}$&  24.85$_{\pm{8.94}}$ & 22.21$_{\pm{9.20}}$ \\
\bottomrule
\label{tab:finetuning}
\end{tabular}
\end{adjustbox}
\end{table*}

\begin{figure*}[htb]
\centering
\includegraphics[width=0.8\textwidth]{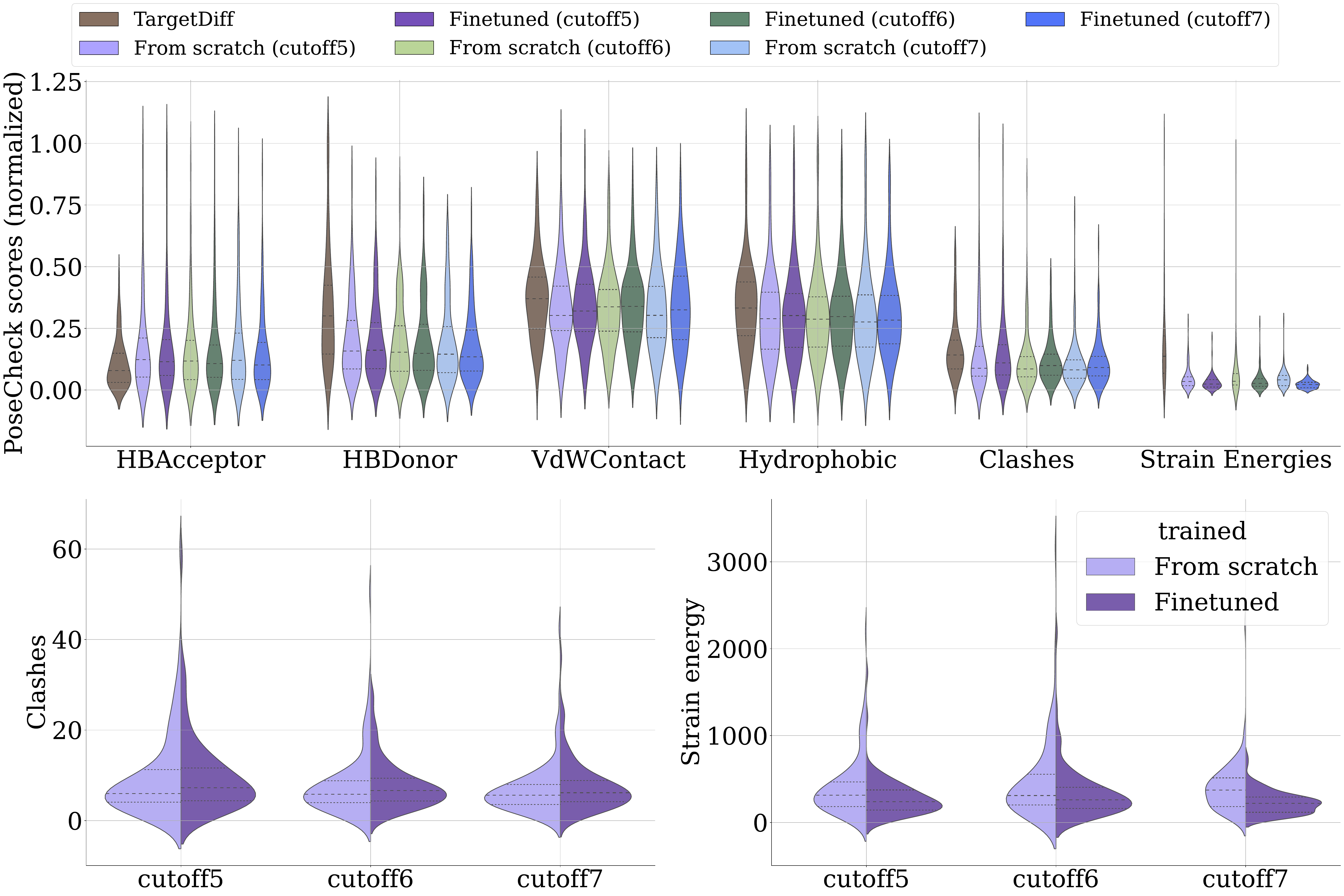}
\caption{The impact of varying dataset cutoffs and employing different training approaches (training from scratch versus pre-training) on the performance of our model and TargetDiff is analyzed. \textbf{Top}: We compare the sample quality using the PoseCheck metrics, where all values are min-max normalized to better evaluate the difference in performance. \textbf{Bottom}: We present the average clash counts (left) and average strain energies (right). Models with lower clashes and strain energies are considered to perform better and are thus preferred.}
\label{fig:posecheck}
\end{figure*}

We extend our evaluation using a range of metrics from PoseCheck \cite{harris2023} to assess their ability to generate ligands that form appropriate poses within the vicinity of the protein pocket. 
However, it is important to clarify that TargetDiff, and \pilot $ $  are not specifically designed or trained to produce exact poses, unlike tools like DiffDock \cite{corso_2023}, which are explicitly developed and trained for docking applications. 
Still, \denovo~models should generate ligand poses  without spatial conflicts, such as clashing with the pocket -- a common issue highlighted in recent studies. \citep{harris2023, buttenschoen2023} 
Furthermore, strain energy is a crucial metric used to evaluate ligands; it measures the energy required to alter a ligand's conformation to fit its binding pose. Those with lower strain energy are generally favorable as they are likely to exhibit stronger binding with the protein. 
Fig.~\ref{fig:posecheck} shows that our pre-trained model significantly excels in terms of reducing strain energy.
Note that the pre-training on molecules without pocket does not lead to an increase of clashes between ligand and pocket atoms in the complex. 
The metrics concerning the number of hydrogen acceptors, donors, van der Waals contacts, and hydrophilicity remain consistent across models.

The reduction in strain energy observed in the pre-trained model might be attributed to two main factors. 
First, the diffusion model is exposed to a vast array of conformers during its pre-training phase, likely featuring low strain energy due to the conformer generation techniques employed. 
This results in the generation of 3D conformers with optimal torsional profiles and minimized torsional strains, contributing to overall lower energy values in the ligands produced. 
Second, the Enamine Real Diversity subset used for pre-training typically includes a wide variety of stable ring systems. 
Thus, the model likely encounters fewer unfavorable ring systems (e.g. 3- or 9-membered rings), which could contribute to higher strain energies.
These insights further underscore the importance of the initial pre-training phase to generate relevant and biologically active ligands, further validating the efficacy of our approach in advancing the field of structure-based drug discovery.

% end WIP

\subsection{Multi-objective de novo generation using importance sampling}

\begin{figure*}[ht]
\centering
  \includegraphics[width=0.8\textwidth]{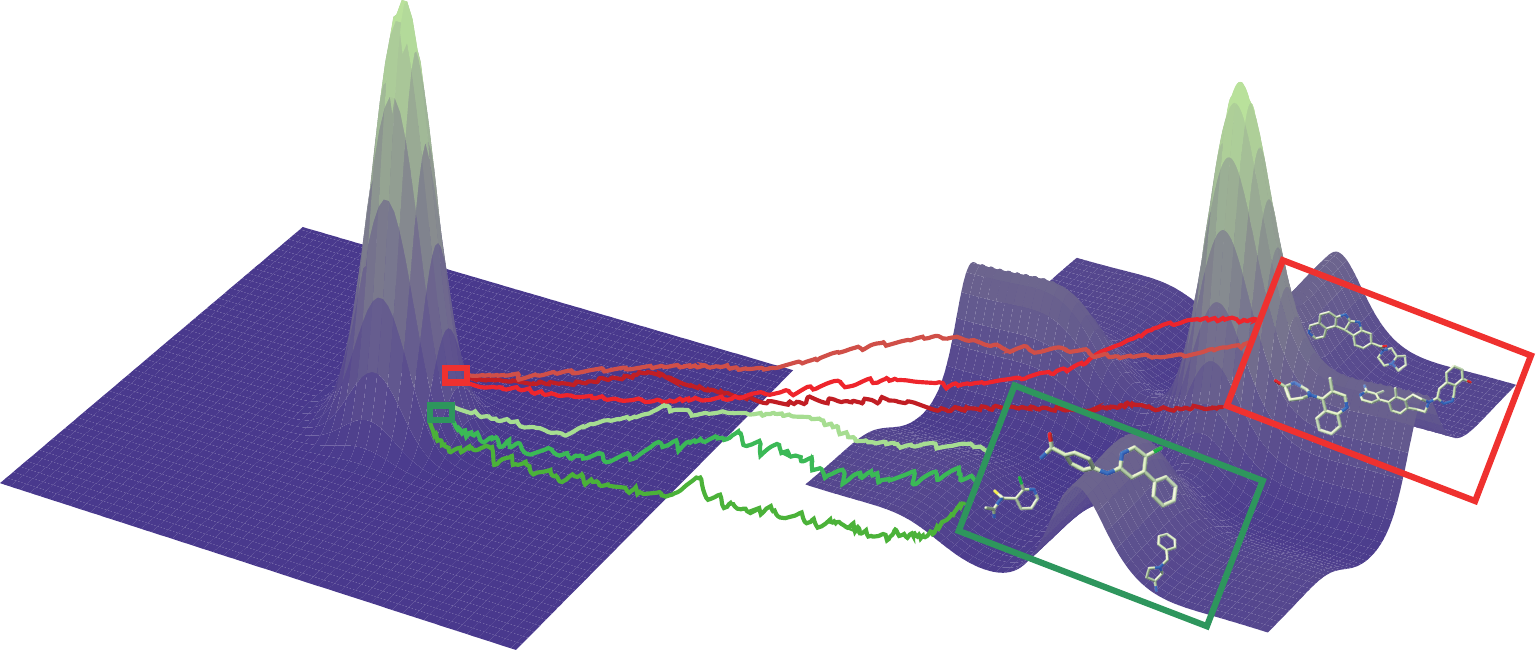}
  \caption{
  Schematical depiction of the importance sampling algorithm. The shape of the prior (left) and target (right) distribution, where ligands at the target distribution are highlighted in two different regions based on a property function, which is synthetic accessibility in this case. At $t=T$ (left), noisy samples are drawn from the prior, and during the reverse trajectory, stochastic paths that lead to promising candidates are selected and de-noised in state-space to converge to samples from the data distribution at $t=0$ (right). Ligands in the green box refer to molecules with high synthetic accessibility according to SA score, while molecules in the red box refer to rather inaccessible ones.}
  \label{fig:importance_sampling}
\end{figure*}

In previous studies utilizing 3D target-aware molecule generation, a significant challenge has been the poor synthetic accessibility (SA) of the generated molecules. These models often produce molecules with complex, fused, and uncommon ring systems, which are difficult to synthesize. \citep{tamgen, guan2023d, schneuing_2023_structurebased} This issue underscores the need for approaches that not only produce molecules with strong binding affinities but also ensure that these molecules can be feasibly synthesized. To address this, we propose a trajectory-based importance sampling method that utilizes property-specific expert models explained in Section \ref{sec:methods}.

The evaluation of the importance sampling approach is performed for both single- and multi-objective optimization scenarios, focusing on SA and docking score guidance. 
We refer to guidance with an SA score model as \textit{SA-conditional} and using a docking score model as \textit{docking-conditional}. 
When both objectives are considered, we refer to the model as \textit{SA-docking-conditional}. 
In each case, the \textit{unconditional} base model is augmented with the respective property model during the sampling process.

\begin{figure*}[htb!]
\centering
\includegraphics[width=0.8\textwidth]{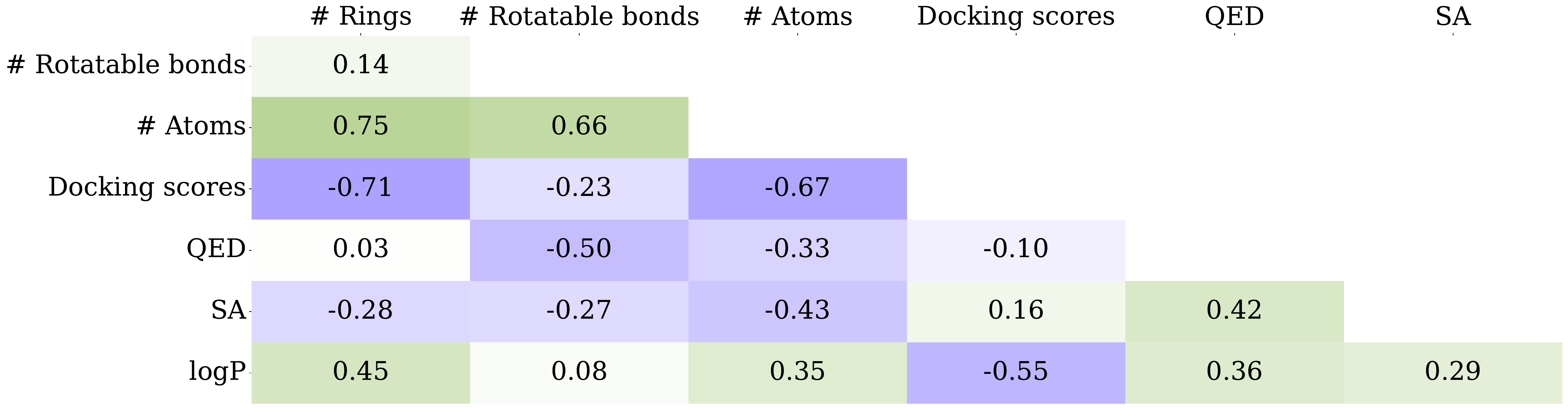}
\caption{Correlation matrix that includes the number of rings, number of atoms, docking scores, quantitative estimate of drug-likeness (QED), and synthetic accessibility (SA) scores for the CrossDocked2020 training set.}
\label{fig:correlation}
\end{figure*}

Table \ref{fig:correlation} shows the correlation matrix of the CrossDocked2020 dataset. 
The SA scores exhibit a negative correlation with ligand size, i.e., larger molecules tend to be less synthetically accessible on average. 
Conversely, the positive correlation between SA scores and QED suggests that molecules with higher QED are generally more synthetically accessible.
Docking scores show a strong negative correlation with both the number of rings and the number of atoms. 
This implies that models driven by docking scores tend to generate larger molecules with more (fused) rings.
However, such molecular characteristics typically result in decreased SA scores and QED, presenting a trade-off between optimizing for docking score and maintaining synthetic feasibility.
By incorporating these insights into our modeling approach, we aim to balance the dual objectives of binding efficacy and synthetic accessibility, thereby enhancing the practical utility of the generated molecules in drug discovery.

Table \ref{tab:docking} shows that our model reproduces the observed correlations of the dataset.
When guiding the \textit{unconditional} model with the SA score, we notice a significant enhancement not only in the SA score, which increases to 0.77, but also improvements in QED and Lipinski's rule of five compliance.
The mean docking scores remain consistent with those of the \textit{unconditional} model.
However, there is a notable reduction of docking performance in the top-10 ligands, consistent with the correlations observed in the dataset.
Conversely, applying docking score guidance exclusively results in diminished SA scores and QED, while the docking scores themselves increase considerably. 
This reflects the trade-offs involved in optimizing for docking efficacy at the expense of synthetic accessibility and drug-likeness.
When applying both SA and docking score guidance, the model achieves comparably high values for SA, QED, and Lipinski, while significantly improving docking scores and outperforming TargetDiff by a large margin. 

To mitigate the adverse impact on SA scores and drug-likeness typically associated with high docking scores of larger molecules, we introduce a normalization strategy where docking scores are adjusted by the square root of the number of atoms per ligand. 
The results of this adjusted model, denoted as \textit{SA-docking-conditional (norm)}, are presented in the last row of Table \ref{tab:docking}. 
Here, we observe a significant increase in docking scores compared to the \textit{unconditional} model, while the SA scores improve to 0.78, compared to 0.77 in the \textit{SA-conditional} model. 
This illustrates how our multi-objective optimization strategy balances different property demands.
Such balanced outcomes are critical for advancing the practical utility of generated molecules in drug discovery, ensuring that they not only bind effectively but are also feasible for synthesis.

\begin{table*}[ht!]
\centering
\caption{Performance comparison among \textit{unconditional} sampling, \textit{SA-conditional}, \textit{docking-conditional}, and \textit{SA-docking-conditional} sampling using the CrossDocked test set, which includes 100 targets. For each target, 100 ligands were sampled. We assessed the performance based on several criteria: mean docking scores obtained from QVina2 re-docking, the top-10 mean docking scores per target, drug-likeness (QED), synthetic accessibility score (SA), compliance with Lipinski's Rule of Five (Lipinski), and mean diversity (Diversity) across targets and ligands.}
\begin{adjustbox}{width=1.0\textwidth,center}
\begin{tabular}{lrrrrrr}\toprule
  Model & QVina2 (All) $\downarrow$ & QVina2 (Top-10\%) $\downarrow$ & QED $\uparrow$ & SA $\uparrow$ & Lipinski $\uparrow$ & Diversity $\uparrow$ \\ \midrule
    Training set & -7.57$_{\pm{2.09}}$ & - & 0.53$_{\pm{0.20}}$ & 0.75$_{\pm{0.10}}$ & 4.57$_{\pm{0.91}}$ & - \\[.3em]
    Test set & -6.88$_{\pm{2.33}}$ & - & 0.47$_{\pm{0.20}}$ & 0.72$_{\pm{0.13}}$ & 4.34$_{\pm{1.14}}$ & - \\[.3em]
    TargetDiff & -7.32$_{\pm{2.47}}$ & -9.67$_{\pm{2.55}}$ & 0.48$_{\pm{0.20}}$ & 0.58$_{\pm{0.13}}$ & 4.59$_{\pm{0.83}}$ & \textbf{0.75}$_{\pm{0.09}}$ \\[.3em]
  %DiffSBDD-cond & -6.95$_{\pm{2.06}}$ & -9.12$_{\pm{2.16}}$ & 0.47$_{\pm{0.21}}$ & 0.58$_{\pm{0.13}}$ & 4.56$_{\pm{0.89}}$ & 0.73$_{\pm{0.07}}$ \\[.3em] %\cline{3-8
    unconditional & -7.33$_{\pm{2.19}}$ & -9.28$_{\pm{2.26}}$ & 0.49$_{\pm{0.22}}$ & 0.64$_{\pm{0.13}}$ & 4.40$_{\pm{1.05}}$ & 0.69$_{\pm{0.07}}$ \\[.3em]
    SA-conditional & -7.32$_{\pm{2.25}}$ & -8.91$_{\pm{2.29}}$ & \textbf{0.58}$_{\pm{0.19}}$ & \textbf{0.77}$_{\pm{0.10}}$ & \textbf{4.82}$_{\pm{0.54}}$ & 0.73$_{\pm{0.08}}$ \\[.3em]
    docking-conditional & \textbf{-9.17}$_{\pm{2.48}}$ & \textbf{-10.94}$_{\pm{2.51}}$ & 0.54$_{\pm{0.13}}$ & 0.62$_{\pm{0.08}}$ & 4.70$_{\pm{0.41}}$ & 0.57$_{\pm{0.10}}$ \\[.3em]
    SA-docking-conditional & -8.35$_{\pm{2.75}}$ & -10.36$_{\pm{2.62}}$ & \textbf{0.58}$_{\pm{0.17}}$ & 0.72$_{\pm{0.12}}$ & \textbf{4.88}$_{\pm{0.44}}$ & 0.68$_{\pm{0.09}}$ \\[.3em]
    SA-docking-conditional (norm) & {-7.92}$_{\pm{2.44}}$ & {-9.85}$_{\pm{2.33}}$ & {0.56}$_{\pm{0.19}}$ & \textbf{0.78}$_{\pm{0.11}}$ & \textbf{4.84}$_{\pm{0.47}}$ & \textbf{0.75}$_{\pm{0.13}}$ \\[.3em]
  \bottomrule
 \label{tab:docking}
\end{tabular}
\end{adjustbox}
\end{table*}

\begin{figure*}[htb!]
\centering
\includegraphics[width=0.8\textwidth]{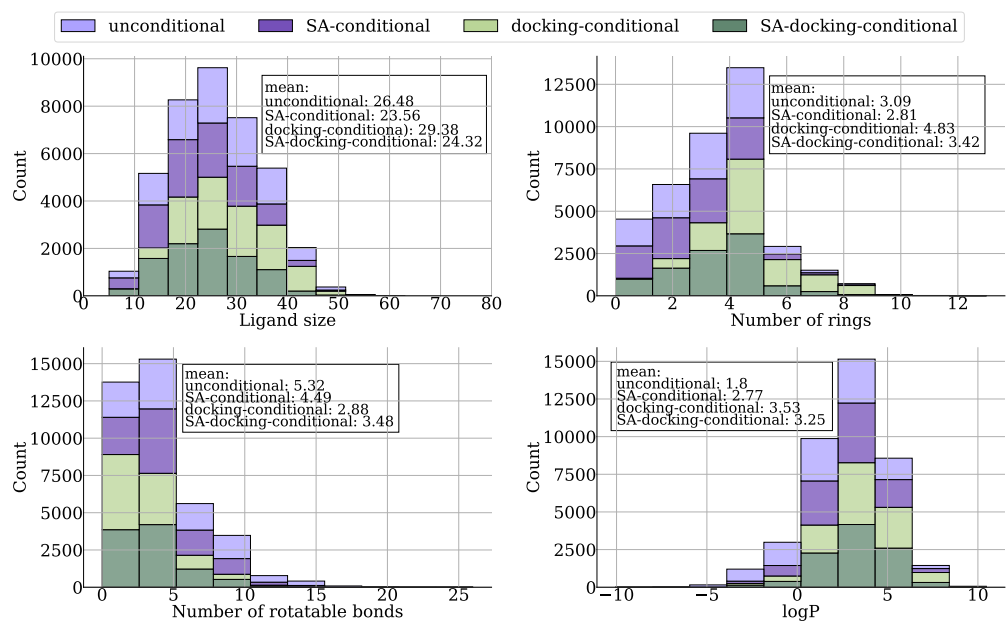}
\caption{Analyze of the distribution of certain ligand characteristics, including size, number of rings, number of rotatable bonds, and logP values, across three sampling methods to show the effect on physicochemical properties: \textit{unconditional} sampling, \textit{SA-conditional}, and SA-docking-conditioned sampling.}
\label{fig:ligand_ring_sizes}
\end{figure*}

We investigate how various molecular properties are affected by the application of guidance to further study the impact of importance sampling guidance on molecular design.
Fig.~\ref{fig:ligand_ring_sizes} shows molecular characteristics such as ligand sizes, number of rings, number of rotatable bonds, and logP values across different models. 
Based on previous observations (Fig.~\ref{fig:correlation}), we expect SA guidance to result in smaller ligands with fewer rings, contrasting with the effect of docking guidance.
First, we determine the most likely ligand size given a target from the training distribution and allow for the addition of up to ten atoms during inference. 
Fig. \ref{fig:ligand_ring_sizes}(top) shows that ligands indeed tend to be smaller and possess fewer rings under SA guidance. 
The \textit{SA-docking-conditional} model, which integrates both SA and docking objectives, represents a balanced compromise between these extremes.

Lipinski's rule of five is an important measure for assessing drug-likeness, including criteria such as the number of rotatable bonds and logP values. 
The number of rotatable bonds exhibits a strong positive correlation with the number of atoms mitigating the slight negative correlation with both SA and docking scores, while logP shows a positive correlation with SA- and docking scores. 
Fig. \ref{fig:ligand_ring_sizes} (bottom) illustrates effective conditioning as both the SA- and \textit{docking-conditional} models generally result in a lower average number of rotatable bonds compared to the \textit{unconditional} model.
In contrast, the partition coefficient logP tends to increase under both conditions.

\begin{figure*}[htb!]
\centering
\includegraphics[width=0.8\textwidth]{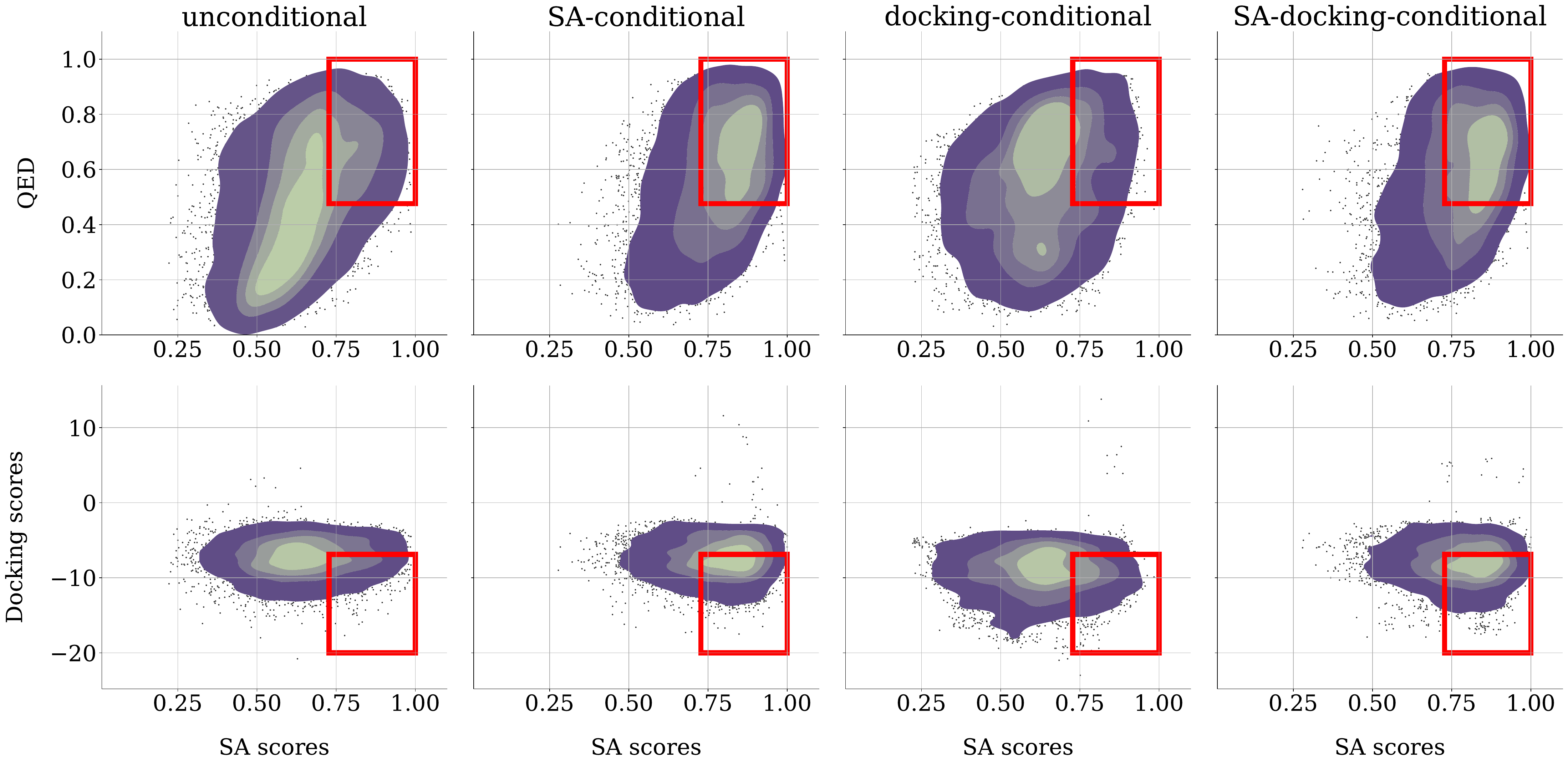}
\caption{Scatter plots with Gaussian kernel density estimation (KDE) were used to illustrate the evolution of QED, SA, and docking scores for all sampled ligands across test targets for different sampling methods: \textit{unconditional}, \textit{SA-conditional}, \textit{docking-conditional}, and \textit{SA-docking-conditional} sampling. Red rectangles within these plots highlight regions where sampled ligands demonstrate superior QED, SA, and docking scores compared to the test set. \textbf{Upper row}: Relationship between QED and SA scores. \textbf{Lower row} Relationship between docking scores and SA scores.}
\label{fig:uncond_vs_cond_gauss}
\end{figure*}

%TODO
Fig.~\ref{fig:uncond_vs_cond_gauss} illustrates the evolution of the sample space across the \textit{unconditional}, \textit{SA-conditional}, \textit{docking-conditional}, and \textit{SA-docking-conditional} models. 
Each plot in this figure includes a red rectangle that identifies the regions where samples exceed the respective means of the test set, indicating improved property scores.
The first row of Fig. \ref{fig:uncond_vs_cond_gauss} compares the drug-likeness (QED) of sampled ligands with their synthetic accessibility (SA) scores.
% The mean QED value in the test set is 0.47, and the mean SA score is 0.72. Thus, the red rectangle encompasses the region where QED values are greater than 0.47 and SA scores exceed 0.72.
The \textit{SA-conditional} model shows a notable shift with most of the sample mass residing within the red rectangle.
Thus, it successfully generates samples with notably higher SA scores compared to both the \textit{unconditional} model and the test set ligands, while largely preserving docking scores.
In contrast, the \textit{docking-conditional} model exhibits lower docking scores on average at the expense of the SA scores.
The \textit{SA-docking-conditional} model demonstrates a good balance, transitioning towards both high SA scores and low docking scores. Remarkably, most of the sampled ligands from this model not only fall within the red rectangle but also significantly surpass the test set ligands in terms of docking scores with equal SA scores as listed in Table \ref{tab:docking}, while the model with normalization improves in both metrics.

Our findings demonstrate that using importance sampling as a guidance mechanism in the diffusion model is a potent strategy for steering the generation of molecules towards desired regions of chemical space. 
The method effectively modifies molecular properties in line with desired multi-objective property profiles, albeit within the constraints set by the data distribution used for training. Unlike classifier-guidance, our approach does not require (prohibitively) expensive backpropagation. Instead, we achieve the aforementioned results using only a few importance sampling steps (forward calls to the surrogate models).

\subsection{Kinodata-3D}

\begin{table*}[ht!]
\centering
\caption{Performance comparison among \textit{unconditional} and \textit{p\ic-conditional} sampling using the Kinodata-3D test set, which includes 10 targets. For each target, 100 ligands were sampled. We assessed the performance based on several criteria: mean docking scores obtained from QVina2 re-docking, the top-10 mean docking scores per target, (predicted) p\ic, drug-likeness (QED), synthetic accessibility score (SA), compliance with Lipinski's Rule of Five (Lipinski), and mean diversity (Diversity) across targets and ligands.}
\begin{adjustbox}{width=1.0\textwidth,center}
\begin{tabular}{lrrrrrrr}\toprule
  Model & Vina (All) $\downarrow$ & Vina (Top-10\%) $\downarrow$ & p\ic $\uparrow$ & QED $\uparrow$ & SA $\uparrow$ & Lipinski $\uparrow$ & Diversity $\uparrow$ \\ \midrule
    Training set & -9.20$_{\pm{1.13}}$ & - & 7.05$_{\pm{1.28}}$ & 0.49$_{\pm{0.16}}$ & 0.75$_{\pm{0.07}}$ & 4.73$_{\pm{0.52}}$ & - \\[.3em]
    Test set & -8.78$_{\pm{1.13}}$ & - & 6.41$_{\pm{1.56}}$ & 0.61$_{\pm{0.14}}$ & 0.79$_{\pm{0.05}}$ & \textbf{4.96}$_{\pm{0.22}}$ & - \\[.3em]
    unconditional & -8.49$_{\pm{1.05}}$ & -9.79$_{\pm{0.87}}$ & 6.28$_{\pm{0.68}}$ &\textbf{0.63}$_{\pm{0.14}}$ & \textbf{0.75}$_{\pm{0.13}}$ & \textbf{4.95}$_{\pm{0.25}}$ & \textbf{0.65}$_{\pm{0.06}}$ \\[.3em]
    p\ic-conditional & -8.60$_{\pm{0.98}}$ & -9.75$_{\pm{0.86}}$ & \textbf{7.65}$_{\pm{0.78}}$ &\textbf{0.62}$_{\pm{0.16}}$ & 0.67$_{\pm{0.09}}$ & \textbf{4.94}$_{\pm{0.28}}$ & 0.57$_{\pm{0.06}}$ \\[.3em]
  \bottomrule
 \label{tab:pic50}
\end{tabular}
\end{adjustbox}
\end{table*}

Kinodata-3D~\cite{Backenköhler_Groß_Wolf_Volkamer_2024} is an \textit{in silico} curated and processed collection of kinase complex cross-docked data designed to facilitate the training of machine learning models on structural protein-ligand complexes with experimental binding affinity data. 
The dataset builds on the cross-docking benchmark established by \citet{Schaller2023.09.11.557138}, adopting a template-based approach. 
In the workflow described by \citet{Backenköhler_Groß_Wolf_Volkamer_2024}, a suitable co-crystallized template complex for each kinase-ligand assay pair is identified from the KLIFS database \cite{10.1093/nar/gkaa895} based on chemical similarity to other co-crystallized ligands of the same kinase. The chosen template is then docked using the Posit software. Kinodata-3D encompasses 140,977 protein-ligand complexes, offering a rich dataset for in-depth study.

Despite the significant advances in virtual screening and the widespread use of docking as a tool for evaluating ligand efficacy, the correlation between docking scores and experimental binding affinities, e.g. measured by the half maximal inhibitory concentration \ic, remains weak at best.
Consequently, reliance on docking scores as stand-ins for binding affinities is potentially misleading.
To address this challenge, we apply our guidance mechanism to directly utilize experimental binding affinities. 
%More accurately, we employ the p\ic values, which are logarithmic transformations of \ic, facilitating easier computational handling and interpretation. 
We leverage the Kinodata-3D dataset, annotated with experimental p\ic~values, to train \pilot $ $  on ligand-kinase complexes.
Simultaneously, we train a property model predicting p\ic, to guide the diffusion model with the proposed importance sampling towards ligands that are more likely to be potent inhibitors. 

We evaluate the models on a hold-out test set comprising ten kinase targets that were not included in either the training or validation datasets. 
The performance of our \textit{p\ic-conditional} model is summarized in Table \ref{tab:pic50}.
The \textit{p\ic-conditional} model shows a significant improvement in predicted mean p\ic values of 7.65$_{\pm{0.78}}$ compared to the test set ligands (6.41$_{\pm{1.56}}$). 
At the same time, it maintains robust performance metrics in terms of docking scores and other critical properties such as QED and compliance with Lipinski's rule of five.
%These outcomes underscore the effectiveness of directly integrating experimental binding data into the model training and guidance processes.

\begin{figure*}[htb]
\centering
\includegraphics[width=0.7\textwidth]{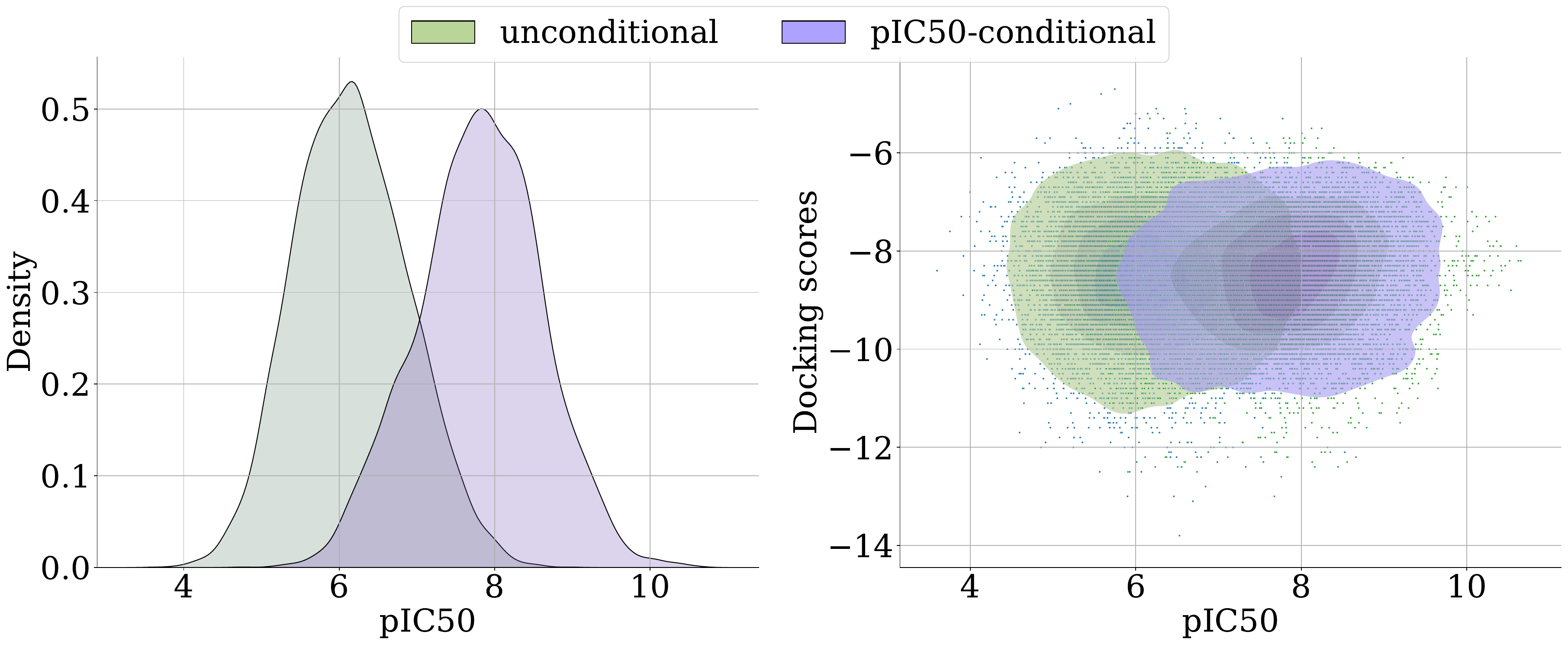}
\caption{\textbf{Left}: Density plot comparing \textit{unconditional} with \textit{p\ic-conditional} sampling. \textbf{Right}: Scatter heatmap overlap of \textit{unconditional} and \textit{p\ic-conditional} samples comparing docking scores and (predicted) p\ic values.}
\label{fig:kinodata_density_ic50}
\end{figure*}

Fig. \ref{fig:kinodata_density_ic50} provides a visual comparison of the sample spaces generated by the \textit{unconditional} and the \textit{p\ic-conditional} model.
We observe a significant shift in the overall density of samples towards higher predicted p\ic~when using the importance sampling guidance (left panel).
Fig. \ref{fig:kinodata_density_ic50} (right) illustrates the relationship between docking scores and p\ic. 
While the \textit{p\ic-conditional} model yields samples with higher p\ic~on average, the ligands maintain competitive docking scores. 
This suggests that the model does not compromise docking efficacy for higher expected p\ic.

\begin{figure*}[htb]
\centering
\includegraphics[width=0.8\textwidth]{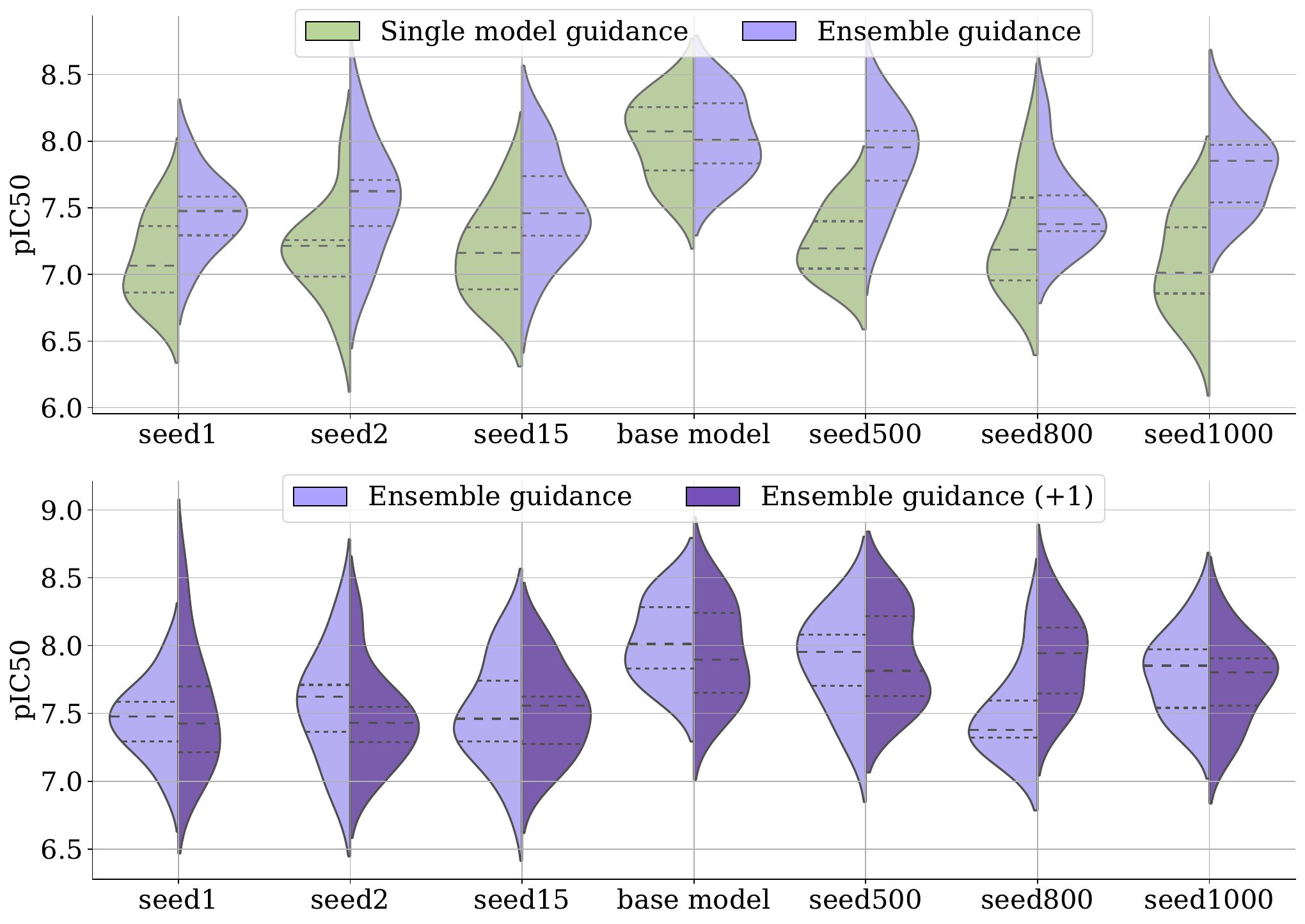}
\caption{A violin plot is used to display the distribution of predicted p\ic values for 100 sampled ligands across ten test set targets, guided either by a single model or an ensemble approach. \textbf{Upper panel}: Ligands generated under single model guidance, where the base model guides itself, or ensemble guidance that includes seed models 500 and 1000. All other models are utilized for evaluating the respective samples. \textbf{Lower panel}: Here, the ensemble guidance for the base model is extended by incorporating an additional model, specifically seed800. This is referred to as "Ensemble guidance (+1)".}
\label{fig:kinodata_ensemble_violin}
\end{figure*}

Note, that the current approach is limited as p\ic~values are inherently noisy, in particular when collected across various data sources.\citep{ic50_noise} 
%Second, experimental validation of these values falls outside the scope of the current study. 
Thus, the predicted binding affinities should be interpreted cautiously. 
To alleviate this problem, we propose to adopt ensemble modeling techniques to enhance the meaningfulness of predictions in the importance sampling pipeline.
Similar approaches are, for example, commonly used for stabilizing machine learning force fields.\citep{ensemble_ff}

Fig. \ref{fig:kinodata_ensemble_violin} (top) demonstrates how ensemble techniques significantly improve the robustness of p\ic~predictions. 
We employ an ensemble of property models for importance sampling guidance. 
Each property model, denoted as seed1, seed2, etc., is trained with a different global seed. 
The base model is used to sample 100 ligands per test target, both with and without ensemble guidance. 
The term \textit{single model guidance} refers to the base model guiding itself. 
We observe that single model guidance results in a notable offset between the predictions of the base model and those of all other property models, indicating poor generalization performance. 
That is, self-guidance exploits the predicted p\ic~value too much, as it was trained on. 
However, with ensemble guidance, even just two additional seed models (seed500 and seed1000) lead to greater improvement in generality.
This enhancement is evident in the p\ic~predictions of all seed models not included in the ensemble guidance (i.e., seed1, seed2, seed15, and seed800).
As shown in Fig. \ref{fig:kinodata_ensemble_violin} (bottom), further increasing the ensemble size, such as by adding another model, here seed800, leads to additional refinement in predictions and consequently, increased generality of p\ic~predictions. 

\section{Conclusions}
We have introduced \pilot, a novel equivariant diffusion-based model tailored for \denovo ~ligand generation conditioned on protein pockets in three-dimensional space. Our research demonstrates the superior performance of \pilot $ $  compared to existing state-of-the-art models in this domain, as evidenced by a comprehensive evaluation across a spectrum of metrics critical in medicinal chemistry and drug design.

A significant finding of our study is the substantial enhancement in downstream performance achieved by pre-training our model on a vast dataset of molecular conformers. This underscores the pivotal role of pre-training in the structure-based drug discovery pipeline, demonstrating its efficacy in improving the quality of generated ligands.
Beyond that, we have proposed a trajectory-based importance sampling strategy, which enables targeted steering of ligand generation towards desired chemical properties. 
This technique guides the generation process towards ligands with desired properties such as synthetic accessibility, drug-likeness, docking scores, and predicted binding affinities by using surrogate models trained on experimental data. This strategy represents an important advancement in structure-based drug discovery, offering researchers a powerful tool to design molecules with tailored properties using 3D equivariant diffusion models.

% While our model exhibits promising capabilities, some potential pitfalls warrant attention. The performance of \pilot heavily relies on the quality and diversity of the training data. Inadequate representation of certain types of protein-ligand interactions could lead to biased or suboptimal ligand generation.

The dependency on the availability and quality of training data remains a critical challenge for deploying AI models like \pilot $ $ in drug discovery pipelines.
In the domain of structure-based drug design, data can often be sparse, noisy, and of varying quality, which significantly impacts the learning and predictive capabilities of ML models.
While our method heavily relies on surrogate models and proxies such as the RDKit synthetic accessibility (SA) scores to estimate the synthesizability of generated ligands, these scores may not fully capture the complexities and practical challenges of medicinal chemistry.
Addressing these challenges will require a concerted effort to enhance data collection practices, improve data quality, and expand the variety of data sources. 

Moving forward, we see potential applications of \pilot $ $ in the drug discovery pipeline by integrating this model with other AI-driven tools and technologies, such as automated synthesis platforms and high-throughput screening to accelerate drug design. Furthermore, the scope of our model may be extended from small molecule drugs to biologic therapeutics involving for example peptides or antibodies. 

\section{Methods}\label{sec:methods}
\subsection{Pocket conditioned 3D diffusion models}
We aim to generate novel molecules \( M \) \denovo, conditioned on a protein pocket \( P \) while optimizing multiple objectives \( c \), such as synthetic accessibility, docking score, and predicted half-maximal inhibitory concentration (\ic). 
Recent developments have utilized 3D diffusion models to implement \( p_\theta(M|P) \), where the task of the model is to denoise an initially random ligand structure, while maintaining the protein pocket as a fixed condition.\citep{schneuing_2023_structurebased, guan2023d, le_cremer2023}
This is achieved by following a stochastic path that targets the distribution of training data, iteratively moving towards more defined structures \( p_\theta(M_{t-1}|M_t, P) \) as illustrated in Figure \ref{fig:graphical-abstract-method}.

During training, the reverse distribution \( p_\theta(M_{t-1}|M_t, P) \) is parameterized using the approach as proposed by \citet{le_cremer2023}.
That is, a noisy ligand \( M_t = (X_t, H_t, E_t) \) at timestep $t$ is represented by perturbed atomic coordinates $X_t$, element types $H_t$, and bond features $E_t$, while the diffusion model \( p_\theta \) is tasked in predicting the noise-free structure \(\hat{M}_0 = (\hat{X}_0, \hat{H}_0, \hat{E}_0)\), acting as denoiser with the inherent goal to iteratively attain a cleaner structure. We optimize the variational lower bound of the log-likelihood $\log p(M_0|P)$ and minimize the timestep-dependent diffusion loss
\begin{equation}
    L_{t} = \frac{1}{2}\left ( w(t) \cdot l_d(M_0,  {p}_\theta(M_t, t, P) \right ),
\end{equation}
where \(l_d:\mathcal{M} \times \mathcal{M} \rightarrow \mathbb{R}^+\) reveals as mean-squared-error loss for 3D coordinates, and cross-entropy loss for discrete-valued data types like atom, bond, and charge-types.\cite{le_cremer2023}
To obtain the noisy ligand \(M_t\), we apply the forward noising process with Gaussian diffusion for continuous valued coordinates, while discrete valued data like atom, bond- and charge-types are perturbed using categorical diffusion which both reads
\begin{align}\label{eq:forward-noising}
    &q({X}_t|{X}_{0}) = \N({X}_{t} | \sqrt{\Bar{\alpha}}_t {X}_0, {(1 - \Bar{\alpha}_t}) {I}) \\
    &q({C}_t|{C}_{0}) = \C({C}_{t} | \Bar{\alpha}_t {C}_0 +(1-\Bar{\alpha}_t)\tilde{{C}} ),
\end{align}
where \(\Bar{\alpha}_t=\prod_{k=1}^t (1-\beta_k) \in (0,1)\) determines a \textit{variance-preserving} (VP) adaptive noise scheduler with empirical distribution \(\tilde{C}\) estimated from the training set for categorical data \((H, E)\).\citep{vignac_2023_midi}

% because the distribution \( M_{t-1} | M_t \) depends on both \( M_t \) and \( \hat{M}_0 \). \kristof{???}
%Specifically, for continuous coordinates, the reverse distribution adheres to a multivariate Gaussian model, while for discrete-valued modalities, it follows a categorical distribution. 
% For further details and technical descriptions we refer to the Appendix.\kristof{methods section?}

\subsection{Multi-objective importance sampling}
\begin{algorithm}[htp!]
    \caption{Importance sampling for property-guided ligand generation, here maximization of $c$} 
    \textbf{Input: }{Pocket $P$, condition $c$, number of ligands $K$, $\tau$ temperature, every importance step $N$, diffusion model $p_\theta$ and property models $p_\delta$.} \\
    \textbf{Output: }{Set of generated ligands $\{M_i\}_{i=1}^K$ conditioned on $(P, c)$.}
    \begin{algorithmic}[1]
        \State Sample $K$ ligands from prior distribution
        $M_T \sim N(0, I) \times C(\hat{p}_c)$
        \For {$t=T-1,\ldots, 1$} \Comment{Run reverse diffusion trajectory}
        \State Sample $M_{t-1} \sim p_\theta(M_{t-1} | M_t, P)$
            \If{$t \mod N = 0$} \Comment{Importance step}
                \For{$k=1, \ldots, K$}
                    \If{optimize for specific $c$}
                        \State $\widetilde{w}_k = p_\delta(c| M_{k,t-1}, P)$ \Comment{Compute probability value}
                    \Else{}     
                        \State $\widetilde{w}_k = f_\delta(M_{k,t-1}, P)$ \Comment{Compute raw property value, here maximization}
                    \EndIf
                \EndFor
                \State Importance weight computation based on population using softmax with temperature $\tau$:
                \State $\{(M_{k, t-1}, \widetilde{w}_k )\}_{k=1}^K$: $~~w_k = \frac{\exp{(\widetilde{w}_k  / \tau)}}{\sum_{j=1}^K \exp{(\widetilde{w}_j / \tau)}}$
                \State Draw new population with replacement:
                \State $\{M_{k, t-1}\}_{k=1}^K \sim \text{Multinomial}(\{M_{k, t-1}\}_{k=1}^K, \{w_k\}_{k=1}^K)$ 
            \EndIf
        \EndFor
        \State return  $\{M_{k, 0}\}_{k=1}^K$
	\end{algorithmic} 
    \label{algorithm:importance-sampling}
\end{algorithm}
To sample ligands from the distribution \( p_\theta(M|P, c) \), we utilize Bayes' theorem to decompose the probability density into \( p_\theta(M|P, c) \propto p_\delta(c|M, P) p_\theta(M|P) \). 
We further assume that multiple properties \( c = (c_1, c_2, \dots, c_k) \) are conditionally independent, leading to the factorization \( p_\delta(c|M, P) = \prod_{l=1}^k p_{\delta_l}(c_l|M, P) \), where each \(p_{\delta_l}(c_l|M, P)\) can be interpreted as an expert surrogate model for a specific property.
These surrogate models must be able to predict the properties of interest at any step of the diffusion trajectory, similar to classifier-guidance. \citep{dhariwal2021diffusion} While classifier-guidance requires backpropagation at every step, making it quickly unfeasible for ligand-pocket complexes with several hundred atoms, our proposed importance sampling approach eliminates the need for backpropagation. Moreover, far fewer steps are needed to update the diffusion model compared to classifier-guidance, which also often tends to steer the model towards adversarial structures. \cite{dhariwal2021diffusion}

As properties such as synthetic accessibility are determined solely based on the ligand, whereas others, like docking scores, depend on the interaction between the ligand and the protein pocket, suitable property predictors \( p_{\delta_i} \) may be defined as required. 
During the sampling process of a set of \( K \) noisy ligands \( \{M_1, M_2, \dots, M_K\} \), we use importance weights derived from \( p_\delta(c|M, P) \) to rank each intermediate noisy sample at its current position in the state space, as described in Algorithm \ref{algorithm:importance-sampling}. Our goal is to generate samples from \(p_\theta(M | c, P) \propto p_\delta(c | M, P) p_\theta(M|P)\) under the condition \(c\), which specifies the property that the ligand \(M\) must achieve. For continuous properties, we choose a Gaussian distribution with a standard deviation of 1 to model \(p(c | M, P)\). Specifically, this takes the form \(p_\delta(c | M, P) = \frac{1}{\sqrt{2\pi}}\exp\left(-\frac{1}{2}(c - f_\delta(M, P))^2\right)\). This formulation also establishes a natural connection to maximum-likelihood training for the property predictor \(f_\delta\).
Since the reverse diffusion trajectory is inherently stochastic, our goal is to preferentially select samples that are most likely to follow a path resulting in ligands meeting the specified conditions \( c \). This process is schematically depicted in Fig. \ref{fig:importance_sampling}.
To accurately predict these conditions, we train \( p_\delta(c|M, P) \) as \( p_\delta(c|M_t, P, t) \) along the forward noising diffusion trajectory, where \( M_t \) represents the state of the ligand at timestep \( t \). 
The property model \( p_\delta \) is trained using the mean squared error and cross-entropy loss for continuous and discrete properties, respectively.
The rationale behind this training approach is that denoising steps closer to the original data distribution retain a clearer signal of the input ligand, making them highly informative.
In contrast, steps closer to the prior noise distribution, although less informative, can still provide valuable discriminative insights for \( p_\delta \). 
This strategy leverages the nuanced progression of information degradation during the diffusion process to efficiently guide the generation of desired ligands without mode collapse.

The algorithm is inspired by the Sequential Monte Carlo (SMC) method.\citep{Doucet2001, sequential_mc_doucet} A similar replacement strategy has previously been applied by \citet{trippe2023diffusion} and \citet{wu2023practical} in the context of diffusion models for protein backbone modeling and motif scaffolding. In Algorithm \ref{algorithm:importance-sampling}, we focus on maximizing property values by scoring each predicted property value among the samples in the population. To achieve this, we employ softmax normalization on the predicted property values \(f_\delta(M_k, P)\) for maximization. If the goal is to minimize a certain property, the predicted property values must be multiplied by \(-1\) to compute the importance weights before applying the softmax operation. These importance weights represent the probability of selecting samples from the finite population set for the next iteration. 
When specific property values \(c\) are desired, instead of relying solely on the predicted property values \(c_k = f_\delta(M_k, P)\), we compute the probability using a Gaussian kernel as described earlier. Notice that we additionally need to employ another normalization scheme to rank each unique probability value. For simplicity, we choose to use softmax normalization again.

% Comment about loss-weighting and in theory the property model would for steps close to prior predict the mean of c to solve the optimization task.
%It's important to note that alternative methods for computing these weights are feasible. 
%For example, if targeting a specific desired property value \( \tilde{c} \), one can utilize an exponential kernel with a predetermined bandwidth \( \sigma_c > 0 \), expressed as \( \tilde{w}_k = \exp\left(-\frac{(p_\delta(M_{k}) - \tilde{c})^2}{\sigma_c}\right)\). This method is analogous to a squared energy function, reflecting the training approach of the property model.\footnote{This reference to a squared energy function is parallel to the method used in training the property model.}
%To convert the set of weights \( \{\tilde{w}_k\}_{k=1}^K \) into a probability distribution among the evaluated particles, these weights are further normalized using the softmax function. This technique ensures that each particle's weight reflects its relative importance in the context of the desired property values, facilitating more effective sampling decisions.

\subsection{Choosing the cutoff for protein-ligand complex creation}\label{sec:cutoff}
The CrossDocked2020 dataset implements a pre-defined cutoff surrounding the bound ligands. For instance, TargetDiff \citep{guan2023d} uses a cutoff region of 10 Ångströms with the centers of mass (CoM) of the residues acting as reference points for measuring distances to ligand atoms. Residues whose CoM are within or equal to the cutoff distance are included in the Protein-Ligand (PL) complex. 
Conversely, DiffSBDD includes the entire residue in the PL complex if any atom within that residue falls inside the cutoff region.\cite{schneuing_2023_structurebased} 
Our work adopts the latter approach as it offers a more physically plausible representation of the interaction space. We ablate different cutoff values \(\{5,6,7\}\angstrom\) on the CrossDocked2020 dataset and observe that the model trained on the 7\AA~cutoff performs best as illustrated in Table \ref{tab:finetuning} for the pre-trained model. We hypothesize that the trade-off between smaller cutoff and model performance is caused by the complexity and tendency to overfit on smaller complexes. Note that a smaller cutoff leads to PL complexes with fewer atoms as shown in Figure and Table B1 in the supplementary materials.
\subsection{Training details} 
We train \pilot $ $ using the network architecture presented in \citet{le_cremer2023} with $T=500$ diffusion timesteps (in contrast to TargetDiff, which uses $T=1000$) with minor modifications to process PL complexes. To perform the message-passing, we compute the interactions on the protein-ligand and the protein-protein graph using a radius-graph with 5 Ångström cutoff. The model comprises 12 message-passing layers and approximately 12.5M parameters.
The Enamine model was pre-trained for 10 epochs with the goal of learning a broad chemical space for ligands without considering any protein pocket context.

We trained the models for 300 epochs from scratch on the CrossDocked2020 dataset. When leveraging the pre-trained Enamine model as a starting point, we only fine-tuned for 100 epochs.

We use the AdamW optimizer with AMSGrad and a learning rate of $2 \cdot 10^{-4}$, weight-decay of $1 \cdot 10^{-12}$, and gradient clipping for values higher than 10 throughout all experiments.

\subsubsection{Property training}
In this work, we utilize a joint training strategy for both the diffusion and property models within a single neural network architecture. Since both models take a noisy ligand \(M_t = (X_t, H_t, E_t)\) as input, the joint model predicts both the clean molecule and the ground-truth property of the input sample, such as synthetic accessibility and/or docking score ($\hat{M}_0$ and $\hat{c}$, respectively). However, importance sampling can be performed using any external model trained on a diffusion trajectory, as long as it uses the same diffusion kernels as the score model. In preliminary studies, we experimented with separately trained models and found that they also worked. However, for simplicity, we used joint training in this work.
%%%%%

% \subsubsection*{Author Contributions}
% If you'd like to, you may include a section for author contributions as is done
% in many journals. This is optional and at the discretion of the authors.

%\subsubsection*{Acknowledgments}
%We acknowledge funding from European Commission grants number 956832 and 101120466 under the Horizon2020 Framework Program for Research and Innovation.

\section*{Acknowledgement}
Julian Cremer received funding from the European Union’s Horizon 2020 research and innovation program under the Marie Skłodowska-Curie Actions grant agreement “Advanced Machine Learning for Innovative Drug Discovery (AIDD)” No. 956832. 

\bibliography{bib}

\providecommand*{\mcitethebibliography}{\thebibliography}
\csname @ifundefined\endcsname{endmcitethebibliography}
{\let\endmcitethebibliography\endthebibliography}{}
\begin{mcitethebibliography}{42}
\providecommand*{\natexlab}[1]{#1}
\providecommand*{\mciteSetBstSublistMode}[1]{}
\providecommand*{\mciteSetBstMaxWidthForm}[2]{}
\providecommand*{\mciteBstWouldAddEndPuncttrue}
  {\def\EndOfBibitem{\unskip.}}
\providecommand*{\mciteBstWouldAddEndPunctfalse}
  {\let\EndOfBibitem\relax}
\providecommand*{\mciteSetBstMidEndSepPunct}[3]{}
\providecommand*{\mciteSetBstSublistLabelBeginEnd}[3]{}
\providecommand*{\EndOfBibitem}{}
\mciteSetBstSublistMode{f}
\mciteSetBstMaxWidthForm{subitem}
{(\emph{\alph{mcitesubitemcount}})}
\mciteSetBstSublistLabelBeginEnd{\mcitemaxwidthsubitemform\space}
{\relax}{\relax}

\bibitem[Anderson(2003)]{Anderson2003}
A.~C. Anderson, \emph{Chemistry \& Biology}, 2003, \textbf{10}, 787--797\relax
\mciteBstWouldAddEndPuncttrue
\mciteSetBstMidEndSepPunct{\mcitedefaultmidpunct}
{\mcitedefaultendpunct}{\mcitedefaultseppunct}\relax
\EndOfBibitem
\bibitem[Batool \emph{et~al.}(2019)Batool, Ahmad, and Choi]{Batool2019}
M.~Batool, B.~Ahmad and S.~Choi, \emph{International Journal of Molecular Sciences}, 2019, \textbf{20}, 2783\relax
\mciteBstWouldAddEndPuncttrue
\mciteSetBstMidEndSepPunct{\mcitedefaultmidpunct}
{\mcitedefaultendpunct}{\mcitedefaultseppunct}\relax
\EndOfBibitem
\bibitem[Green \emph{et~al.}(2021)Green, Koes, and Durrant]{deepfrag_2021}
H.~Green, D.~R. Koes and J.~D. Durrant, \emph{Chem. Sci.}, 2021, \textbf{12}, 8036--8047\relax
\mciteBstWouldAddEndPuncttrue
\mciteSetBstMidEndSepPunct{\mcitedefaultmidpunct}
{\mcitedefaultendpunct}{\mcitedefaultseppunct}\relax
\EndOfBibitem
\bibitem[Luo \emph{et~al.}(2021)Luo, Guan, Ma, and Peng]{luo_2021}
S.~Luo, J.~Guan, J.~Ma and J.~Peng, Advances in Neural Information Processing Systems, 2021, pp. 6229--6239\relax
\mciteBstWouldAddEndPuncttrue
\mciteSetBstMidEndSepPunct{\mcitedefaultmidpunct}
{\mcitedefaultendpunct}{\mcitedefaultseppunct}\relax
\EndOfBibitem
\bibitem[Ragoza \emph{et~al.}(2022)Ragoza, Masuda, and Koes]{Ragoza2022}
M.~Ragoza, T.~Masuda and D.~R. Koes, \emph{Chem. Sci.}, 2022, \textbf{13}, 2701--2713\relax
\mciteBstWouldAddEndPuncttrue
\mciteSetBstMidEndSepPunct{\mcitedefaultmidpunct}
{\mcitedefaultendpunct}{\mcitedefaultseppunct}\relax
\EndOfBibitem
\bibitem[Liu \emph{et~al.}(2022)Liu, Luo, Uchino, Maruhashi, and Ji]{liu2022}
M.~Liu, Y.~Luo, K.~Uchino, K.~Maruhashi and S.~Ji, Proceedings of the 39th International Conference on Machine Learning, 2022, pp. 13912--13924\relax
\mciteBstWouldAddEndPuncttrue
\mciteSetBstMidEndSepPunct{\mcitedefaultmidpunct}
{\mcitedefaultendpunct}{\mcitedefaultseppunct}\relax
\EndOfBibitem
\bibitem[Tan \emph{et~al.}(2022)Tan, Gao, and Li]{tan2022}
C.~Tan, Z.~Gao and S.~Z. Li, \emph{Target-aware Molecular Graph Generation}, 2022, \url{https://arxiv.org/abs/2202.04829}\relax
\mciteBstWouldAddEndPuncttrue
\mciteSetBstMidEndSepPunct{\mcitedefaultmidpunct}
{\mcitedefaultendpunct}{\mcitedefaultseppunct}\relax
\EndOfBibitem
\bibitem[Peng \emph{et~al.}(2022)Peng, Luo, Guan, Xie, Peng, and Ma]{peng_2022}
X.~Peng, S.~Luo, J.~Guan, Q.~Xie, J.~Peng and J.~Ma, Proceedings of the 39th International Conference on Machine Learning, 2022, pp. 17644--17655\relax
\mciteBstWouldAddEndPuncttrue
\mciteSetBstMidEndSepPunct{\mcitedefaultmidpunct}
{\mcitedefaultendpunct}{\mcitedefaultseppunct}\relax
\EndOfBibitem
\bibitem[Powers \emph{et~al.}(2023)Powers, Yu, Suriana, Koodli, Lu, Paggi, and Dror]{dror_2023}
A.~S. Powers, H.~H. Yu, P.~Suriana, R.~V. Koodli, T.~Lu, J.~M. Paggi and R.~O. Dror, \emph{ACS Central Science}, 2023, \textbf{9}, 2257--2267\relax
\mciteBstWouldAddEndPuncttrue
\mciteSetBstMidEndSepPunct{\mcitedefaultmidpunct}
{\mcitedefaultendpunct}{\mcitedefaultseppunct}\relax
\EndOfBibitem
\bibitem[Guan \emph{et~al.}(2023)Guan, Qian, Peng, Su, Peng, and Ma]{guan2023d}
J.~Guan, W.~W. Qian, X.~Peng, Y.~Su, J.~Peng and J.~Ma, The Eleventh International Conference on Learning Representations, 2023\relax
\mciteBstWouldAddEndPuncttrue
\mciteSetBstMidEndSepPunct{\mcitedefaultmidpunct}
{\mcitedefaultendpunct}{\mcitedefaultseppunct}\relax
\EndOfBibitem
\bibitem[Schneuing \emph{et~al.}(2023)Schneuing, Du, Harris, Jamasb, Igashov, Du, Blundell, Lió, Gomes, Welling, Bronstein, and Correia]{schneuing_2023_structurebased}
A.~Schneuing, Y.~Du, C.~Harris, A.~Jamasb, I.~Igashov, W.~Du, T.~Blundell, P.~Lió, C.~Gomes, M.~Welling, M.~Bronstein and B.~Correia, \emph{Structure-based Drug Design with Equivariant Diffusion Models}, 2023, \url{https://arxiv.org/abs/2210.13695}\relax
\mciteBstWouldAddEndPuncttrue
\mciteSetBstMidEndSepPunct{\mcitedefaultmidpunct}
{\mcitedefaultendpunct}{\mcitedefaultseppunct}\relax
\EndOfBibitem
\bibitem[Corso \emph{et~al.}(2023)Corso, St{\"a}rk, Jing, Barzilay, and Jaakkola]{corso_2023}
G.~Corso, H.~St{\"a}rk, B.~Jing, R.~Barzilay and T.~S. Jaakkola, The Eleventh International Conference on Learning Representations, 2023\relax
\mciteBstWouldAddEndPuncttrue
\mciteSetBstMidEndSepPunct{\mcitedefaultmidpunct}
{\mcitedefaultendpunct}{\mcitedefaultseppunct}\relax
\EndOfBibitem
\bibitem[Zhu \emph{et~al.}(2024)Zhu, Gu, Pei, and Lai]{luhua_2024}
J.~Zhu, Z.~Gu, J.~Pei and L.~Lai, \emph{Chem. Sci.}, 2024,  --\relax
\mciteBstWouldAddEndPuncttrue
\mciteSetBstMidEndSepPunct{\mcitedefaultmidpunct}
{\mcitedefaultendpunct}{\mcitedefaultseppunct}\relax
\EndOfBibitem
\bibitem[Xia \emph{et~al.}(2024)Xia, Wu, Deng, Liu, Zhang, Guo, Cui, Pei, Wu, Xie, Chen, Lu, Hu, Wu, Chan, Chen, Zhou, Yu, Liu, Guo, Qin, and Liu]{tamgen}
Y.~Xia, K.~Wu, P.~Deng, R.~Liu, Y.~Zhang, H.~Guo, Y.~Cui, Q.~Pei, L.~Wu, S.~Xie, S.~Chen, X.~Lu, S.~Hu, J.~Wu, C.-K. Chan, S.~Chen, L.~Zhou, N.~Yu, H.~Liu, J.~Guo, T.~Qin and T.-Y. Liu, \emph{Target-aware Molecule Generation for Drug Design Using a Chemical Language Model}, 2024, \url{https://www.biorxiv.org/content/early/2024/01/08/2024.01.08.574635}\relax
\mciteBstWouldAddEndPuncttrue
\mciteSetBstMidEndSepPunct{\mcitedefaultmidpunct}
{\mcitedefaultendpunct}{\mcitedefaultseppunct}\relax
\EndOfBibitem
\bibitem[Gómez-Bombarelli \emph{et~al.}(2018)Gómez-Bombarelli, Wei, Duvenaud, Hernández-Lobato, Sánchez-Lengeling, Sheberla, Aguilera-Iparraguirre, Hirzel, Adams, and Aspuru-Guzik]{bombarelli_vae}
R.~Gómez-Bombarelli, J.~N. Wei, D.~Duvenaud, J.~M. Hernández-Lobato, B.~Sánchez-Lengeling, D.~Sheberla, J.~Aguilera-Iparraguirre, T.~D. Hirzel, R.~P. Adams and A.~Aspuru-Guzik, \emph{ACS Central Science}, 2018, \textbf{4}, 268--276\relax
\mciteBstWouldAddEndPuncttrue
\mciteSetBstMidEndSepPunct{\mcitedefaultmidpunct}
{\mcitedefaultendpunct}{\mcitedefaultseppunct}\relax
\EndOfBibitem
\bibitem[Winter \emph{et~al.}(2019)Winter, Montanari, Steffen, Briem, Noé, and Clevert]{winter_mso_2019}
R.~Winter, F.~Montanari, A.~Steffen, H.~Briem, F.~Noé and D.-A. Clevert, \emph{Chem. Sci.}, 2019, \textbf{10}, 8016--8024\relax
\mciteBstWouldAddEndPuncttrue
\mciteSetBstMidEndSepPunct{\mcitedefaultmidpunct}
{\mcitedefaultendpunct}{\mcitedefaultseppunct}\relax
\EndOfBibitem
\bibitem[Dhariwal and Nichol(2021)]{dhariwal2021diffusion}
P.~Dhariwal and A.~Q. Nichol, Advances in Neural Information Processing Systems, 2021\relax
\mciteBstWouldAddEndPuncttrue
\mciteSetBstMidEndSepPunct{\mcitedefaultmidpunct}
{\mcitedefaultendpunct}{\mcitedefaultseppunct}\relax
\EndOfBibitem
\bibitem[Sterling and Irwin(2015)]{irwin_zinc15}
T.~Sterling and J.~J. Irwin, \emph{Journal of Chemical Information and Modeling}, 2015, \textbf{55}, 2324--2337\relax
\mciteBstWouldAddEndPuncttrue
\mciteSetBstMidEndSepPunct{\mcitedefaultmidpunct}
{\mcitedefaultendpunct}{\mcitedefaultseppunct}\relax
\EndOfBibitem
\bibitem[Hawkins and Nicholls(2012)]{hawkins_2011_oeomega}
P.~C.~D. Hawkins and A.~Nicholls, \emph{Journal of Chemical Information and Modeling}, 2012, \textbf{52}, 2919--2936\relax
\mciteBstWouldAddEndPuncttrue
\mciteSetBstMidEndSepPunct{\mcitedefaultmidpunct}
{\mcitedefaultendpunct}{\mcitedefaultseppunct}\relax
\EndOfBibitem
\bibitem[Francoeur \emph{et~al.}(2020)Francoeur, Masuda, Sunseri, Jia, Iovanisci, Snyder, and Koes]{crossdocked}
P.~G. Francoeur, T.~Masuda, J.~Sunseri, A.~Jia, R.~B. Iovanisci, I.~Snyder and D.~R. Koes, \emph{Journal of Chemical Information and Modeling}, 2020, \textbf{60}, 4200--4215\relax
\mciteBstWouldAddEndPuncttrue
\mciteSetBstMidEndSepPunct{\mcitedefaultmidpunct}
{\mcitedefaultendpunct}{\mcitedefaultseppunct}\relax
\EndOfBibitem
\bibitem[Brown \emph{et~al.}(2020)Brown, Mann, Ryder, Subbiah, Kaplan, Dhariwal, Neelakantan, Shyam, Sastry, Askell, Agarwal, Herbert-Voss, Krueger, Henighan, Child, Ramesh, Ziegler, Wu, Winter, Hesse, Chen, Sigler, Litwin, Gray, Chess, Clark, Berner, McCandlish, Radford, Sutskever, and Amodei]{gpt3}
T.~Brown, B.~Mann, N.~Ryder, M.~Subbiah, J.~D. Kaplan, P.~Dhariwal, A.~Neelakantan, P.~Shyam, G.~Sastry, A.~Askell, S.~Agarwal, A.~Herbert-Voss, G.~Krueger, T.~Henighan, R.~Child, A.~Ramesh, D.~Ziegler, J.~Wu, C.~Winter, C.~Hesse, M.~Chen, E.~Sigler, M.~Litwin, S.~Gray, B.~Chess, J.~Clark, C.~Berner, S.~McCandlish, A.~Radford, I.~Sutskever and D.~Amodei, Advances in Neural Information Processing Systems, 2020, pp. 1877--1901\relax
\mciteBstWouldAddEndPuncttrue
\mciteSetBstMidEndSepPunct{\mcitedefaultmidpunct}
{\mcitedefaultendpunct}{\mcitedefaultseppunct}\relax
\EndOfBibitem
\bibitem[Devlin \emph{et~al.}(2018)Devlin, Chang, Lee, and Toutanova]{bert}
J.~Devlin, M.-W. Chang, K.~Lee and K.~Toutanova, \emph{BERT: Pre-training of Deep Bidirectional Transformers for Language Understanding}, 2018, \url{https://arxiv.org/abs/1810.04805}\relax
\mciteBstWouldAddEndPuncttrue
\mciteSetBstMidEndSepPunct{\mcitedefaultmidpunct}
{\mcitedefaultendpunct}{\mcitedefaultseppunct}\relax
\EndOfBibitem
\bibitem[Winter \emph{et~al.}(2019)Winter, Montanari, Noé, and Clevert]{winter_cddd_2019}
R.~Winter, F.~Montanari, F.~Noé and D.-A. Clevert, \emph{Chem. Sci.}, 2019, \textbf{10}, 1692--1701\relax
\mciteBstWouldAddEndPuncttrue
\mciteSetBstMidEndSepPunct{\mcitedefaultmidpunct}
{\mcitedefaultendpunct}{\mcitedefaultseppunct}\relax
\EndOfBibitem
\bibitem[Liu \emph{et~al.}(2023)Liu, Guo, and Tang]{liu2023molecular}
S.~Liu, H.~Guo and J.~Tang, The Eleventh International Conference on Learning Representations, 2023\relax
\mciteBstWouldAddEndPuncttrue
\mciteSetBstMidEndSepPunct{\mcitedefaultmidpunct}
{\mcitedefaultendpunct}{\mcitedefaultseppunct}\relax
\EndOfBibitem
\bibitem[Zaidi \emph{et~al.}(2023)Zaidi, Schaarschmidt, Martens, Kim, Teh, Sanchez-Gonzalez, Battaglia, Pascanu, and Godwin]{zaidi2023pretraining}
S.~Zaidi, M.~Schaarschmidt, J.~Martens, H.~Kim, Y.~W. Teh, A.~Sanchez-Gonzalez, P.~Battaglia, R.~Pascanu and J.~Godwin, The Eleventh International Conference on Learning Representations, 2023\relax
\mciteBstWouldAddEndPuncttrue
\mciteSetBstMidEndSepPunct{\mcitedefaultmidpunct}
{\mcitedefaultendpunct}{\mcitedefaultseppunct}\relax
\EndOfBibitem
\bibitem[Le \emph{et~al.}(2024)Le, Cremer, Noé, Clevert, and Schütt]{le_cremer2023}
T.~Le, J.~Cremer, F.~Noé, D.-A. Clevert and K.~Schütt, The Twelfth International Conference on Learning Representations, 2024\relax
\mciteBstWouldAddEndPuncttrue
\mciteSetBstMidEndSepPunct{\mcitedefaultmidpunct}
{\mcitedefaultendpunct}{\mcitedefaultseppunct}\relax
\EndOfBibitem
\bibitem[Buttenschoen \emph{et~al.}(2024)Buttenschoen, Morris, and Deane]{buttenschoen2023}
M.~Buttenschoen, G.~M. Morris and C.~M. Deane, \emph{PoseBusters: AI-based docking methods fail to generate physically valid poses or generalise to novel sequences}, 2024, \url{http://dx.doi.org/10.1039/D3SC04185A}\relax
\mciteBstWouldAddEndPuncttrue
\mciteSetBstMidEndSepPunct{\mcitedefaultmidpunct}
{\mcitedefaultendpunct}{\mcitedefaultseppunct}\relax
\EndOfBibitem
\bibitem[Harris \emph{et~al.}(2023)Harris, Didi, Jamasb, Joshi, Mathis, Lio, and Blundell]{harris2023}
C.~Harris, K.~Didi, A.~R. Jamasb, C.~K. Joshi, S.~V. Mathis, P.~Lio and T.~Blundell, \emph{Benchmarking Generated Poses: How Rational is Structure-based Drug Design with Generative Models?}, 2023\relax
\mciteBstWouldAddEndPuncttrue
\mciteSetBstMidEndSepPunct{\mcitedefaultmidpunct}
{\mcitedefaultendpunct}{\mcitedefaultseppunct}\relax
\EndOfBibitem
\bibitem[Backenköhler \emph{et~al.}(2024)Backenköhler, Groß, Wolf, and Volkamer]{Backenköhler_Groß_Wolf_Volkamer_2024}
M.~Backenköhler, J.~Groß, V.~Wolf and A.~Volkamer, \emph{Guided docking as a data generation approach facilitates structure-based machine learning on kinases}, 2024\relax
\mciteBstWouldAddEndPuncttrue
\mciteSetBstMidEndSepPunct{\mcitedefaultmidpunct}
{\mcitedefaultendpunct}{\mcitedefaultseppunct}\relax
\EndOfBibitem
\bibitem[Schaller \emph{et~al.}(2023)Schaller, Christ, Chodera, and Volkamer]{Schaller2023.09.11.557138}
D.~Schaller, C.~D. Christ, J.~D. Chodera and A.~Volkamer, \emph{Benchmarking Cross-Docking Strategies for Structure-Informed Machine Learning in Kinase Drug Discovery}, 2023, \url{https://www.biorxiv.org/content/early/2023/09/14/2023.09.11.557138}\relax
\mciteBstWouldAddEndPuncttrue
\mciteSetBstMidEndSepPunct{\mcitedefaultmidpunct}
{\mcitedefaultendpunct}{\mcitedefaultseppunct}\relax
\EndOfBibitem
\bibitem[Kanev \emph{et~al.}(2020)Kanev, de~Graaf, Westerman, de~Esch, and Kooistra]{10.1093/nar/gkaa895}
G.~K. Kanev, C.~de~Graaf, B.~A. Westerman, I.~J.~P. de~Esch and A.~J. Kooistra, \emph{Nucleic Acids Research}, 2020, \textbf{49}, D562--D569\relax
\mciteBstWouldAddEndPuncttrue
\mciteSetBstMidEndSepPunct{\mcitedefaultmidpunct}
{\mcitedefaultendpunct}{\mcitedefaultseppunct}\relax
\EndOfBibitem
\bibitem[Landrum and Riniker(2024)]{ic50_noise}
G.~A. Landrum and S.~Riniker, \emph{Journal of Chemical Information and Modeling}, 2024, \textbf{64}, 1560--1567\relax
\mciteBstWouldAddEndPuncttrue
\mciteSetBstMidEndSepPunct{\mcitedefaultmidpunct}
{\mcitedefaultendpunct}{\mcitedefaultseppunct}\relax
\EndOfBibitem
\bibitem[Unke \emph{et~al.}(2021)Unke, Chmiela, Sauceda, Gastegger, Poltavsky, Sch{\"u}tt, Tkatchenko, and M{\"u}ller]{ensemble_ff}
O.~T. Unke, S.~Chmiela, H.~E. Sauceda, M.~Gastegger, I.~Poltavsky, K.~T. Sch{\"u}tt, A.~Tkatchenko and K.-R. M{\"u}ller, \emph{Chem. Rev.}, 2021, \textbf{121}, 10142--10186\relax
\mciteBstWouldAddEndPuncttrue
\mciteSetBstMidEndSepPunct{\mcitedefaultmidpunct}
{\mcitedefaultendpunct}{\mcitedefaultseppunct}\relax
\EndOfBibitem
\bibitem[Vignac \emph{et~al.}(2023)Vignac, Osman, Toni, and Frossard]{vignac_2023_midi}
C.~Vignac, N.~Osman, L.~Toni and P.~Frossard, Machine Learning and Knowledge Discovery in Databases: Research Track - European Conference, {ECML} {PKDD} 2023, Turin, Italy, September 18-22, 2023, Proceedings, Part {II}, 2023, pp. 560--576\relax
\mciteBstWouldAddEndPuncttrue
\mciteSetBstMidEndSepPunct{\mcitedefaultmidpunct}
{\mcitedefaultendpunct}{\mcitedefaultseppunct}\relax
\EndOfBibitem
\bibitem[Doucet \emph{et~al.}(2001)Doucet, de~Freitas, and Gordon]{Doucet2001}
A.~Doucet, N.~de~Freitas and N.~Gordon, in \emph{An Introduction to Sequential Monte Carlo Methods}, Springer New York, New York, NY, 2001, pp. 3--14\relax
\mciteBstWouldAddEndPuncttrue
\mciteSetBstMidEndSepPunct{\mcitedefaultmidpunct}
{\mcitedefaultendpunct}{\mcitedefaultseppunct}\relax
\EndOfBibitem
\bibitem[Del~Moral \emph{et~al.}(2006)Del~Moral, Doucet, and Jasra]{sequential_mc_doucet}
P.~Del~Moral, A.~Doucet and A.~Jasra, \emph{Journal of the Royal Statistical Society Series B: Statistical Methodology}, 2006, \textbf{68}, 411--436\relax
\mciteBstWouldAddEndPuncttrue
\mciteSetBstMidEndSepPunct{\mcitedefaultmidpunct}
{\mcitedefaultendpunct}{\mcitedefaultseppunct}\relax
\EndOfBibitem
\bibitem[Trippe \emph{et~al.}(2023)Trippe, Yim, Tischer, Baker, Broderick, Barzilay, and Jaakkola]{trippe2023diffusion}
B.~L. Trippe, J.~Yim, D.~Tischer, D.~Baker, T.~Broderick, R.~Barzilay and T.~S. Jaakkola, The Eleventh International Conference on Learning Representations, 2023\relax
\mciteBstWouldAddEndPuncttrue
\mciteSetBstMidEndSepPunct{\mcitedefaultmidpunct}
{\mcitedefaultendpunct}{\mcitedefaultseppunct}\relax
\EndOfBibitem
\bibitem[Wu \emph{et~al.}(2023)Wu, Trippe, Naesseth, Cunningham, and Blei]{wu2023practical}
L.~Wu, B.~L. Trippe, C.~A. Naesseth, J.~P. Cunningham and D.~Blei, Thirty-seventh Conference on Neural Information Processing Systems, 2023\relax
\mciteBstWouldAddEndPuncttrue
\mciteSetBstMidEndSepPunct{\mcitedefaultmidpunct}
{\mcitedefaultendpunct}{\mcitedefaultseppunct}\relax
\EndOfBibitem
\bibitem[Taylor \emph{et~al.}(2014)Taylor, MacCoss, and Lawson]{rings_in_drugs}
R.~D. Taylor, M.~MacCoss and A.~D.~G. Lawson, \emph{Journal of Medicinal Chemistry}, 2014, \textbf{57}, 5845--5859\relax
\mciteBstWouldAddEndPuncttrue
\mciteSetBstMidEndSepPunct{\mcitedefaultmidpunct}
{\mcitedefaultendpunct}{\mcitedefaultseppunct}\relax
\EndOfBibitem
\bibitem[Yu \emph{et~al.}(2022)Yu, Zhang, Wang, Li, Li, and Wei]{7membered}
X.-C. Yu, C.-C. Zhang, L.-T. Wang, J.-Z. Li, T.~Li and W.-T. Wei, \emph{Organic Chemistry Frontiers}, 2022, \textbf{9}, 4757--4781\relax
\mciteBstWouldAddEndPuncttrue
\mciteSetBstMidEndSepPunct{\mcitedefaultmidpunct}
{\mcitedefaultendpunct}{\mcitedefaultseppunct}\relax
\EndOfBibitem
\bibitem[Rusu \emph{et~al.}(2023)Rusu, Moga, Uncu, and Hancu]{heterocycles}
A.~Rusu, I.-M. Moga, L.~Uncu and G.~Hancu, \emph{Pharmaceutics}, 2023, \textbf{15}, 2\relax
\mciteBstWouldAddEndPuncttrue
\mciteSetBstMidEndSepPunct{\mcitedefaultmidpunct}
{\mcitedefaultendpunct}{\mcitedefaultseppunct}\relax
\EndOfBibitem
\bibitem[Jampilek(2019)]{heterocycles_2}
J.~Jampilek, \emph{Molecules}, 2019, \textbf{24}, 6\relax
\mciteBstWouldAddEndPuncttrue
\mciteSetBstMidEndSepPunct{\mcitedefaultmidpunct}
{\mcitedefaultendpunct}{\mcitedefaultseppunct}\relax
\EndOfBibitem
\end{mcitethebibliography}
\bibliographystyle{rsc}

\newpage
\appendix
\FloatBarrier

\section{Learning Curves: From Scratch vs Fine-tuned}

\begin{figure}[htb!]
\centering
\includegraphics[width=0.75\textwidth]{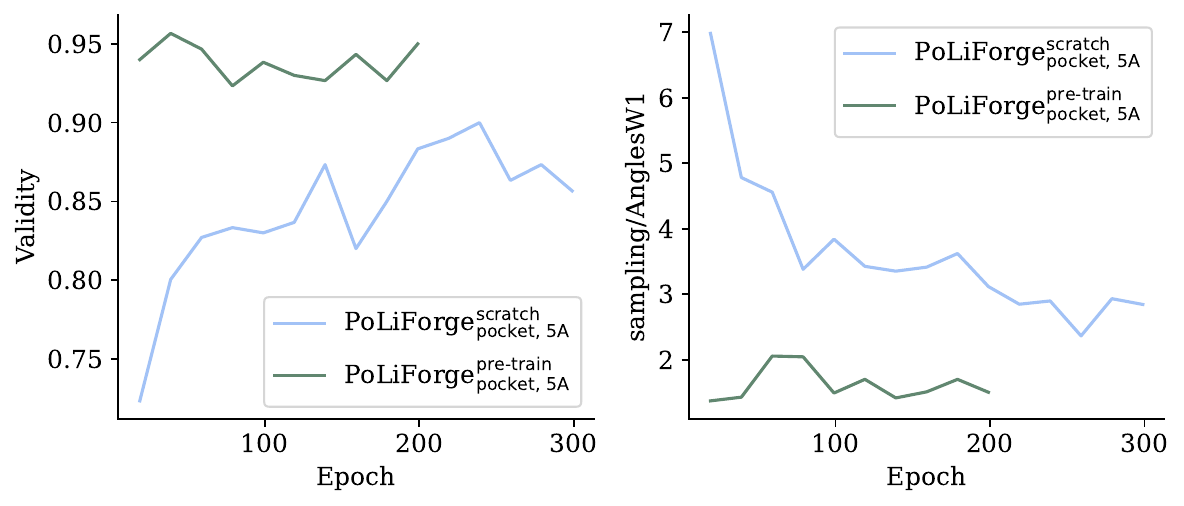}
\caption{Learning curves comparison for the from scratch trained model against fine-tuned model. For better visibility, we fine-tuned the pre-trained model for 200 epochs to obtain more metrics for visualisation.
We show the molecule validity as well as samples/AnglesW1 metric and observe that the fine-tuned model achieves much better metrics already after 20 epochs of training.}
\label{fig:fs-ft-learning_curves.pdf}
\end{figure}
We observe that the pre-trained \pilot~model achieves faster training convergence compared to the model that is trained from scratch on the CrossDocked dataset. In Figure \ref{fig:fs-ft-learning_curves.pdf} we show the evaluation curves for both models trained on the 5A PL-complex dataset, i.e., comparing  \pilot$_\text{pocket, 5A}^{\text{scratch}}$ against \pilot$_\text{pocket, 5A}^{\text{pre-train}}$. As shown, the pre-trained model achieves superior metrics compared to the model trained from scratch already in the first evaluation period after 20 epochs of training. Specifically, the molecule validity for the fine-tuned model accomplishes a highest value of $95.67$\% after 40 epochs while the model trained from scratch only achieves a maximum molecule validity of $90.00$\% after 240 epochs of training. As the pre-trained model has learned on a vast chemical space from the Enamine Real Diversity, this model achieved to learn general chemistry rules related to valency. Nonetheless, when trained on CrossDocked an extensive distribution shift is expected, since in this scenario, the model inputs a much larger sample in form of a protein-ligand complex. Learning correct geometries how a ligand might fit into a protein pocket is a non-trivial task. Although both models are not explicitly trained to generate a ligand that fit a pocket in a physical sense, like a docking tool, the ligands generated by the fine-tuned model achieves predominant angles distribution metrics compared to the from scratch model. This means that the ligands sampled from the fine-tuned model attain better geometries resembles the angles statistics present in protein-ligand-complexes in the validation set.

\section{CrossDocked2020: Analysis}

\subsection{Pocket Size Distribution}

\begin{figure}[htb!]
\centering
\includegraphics[width=1.\textwidth]{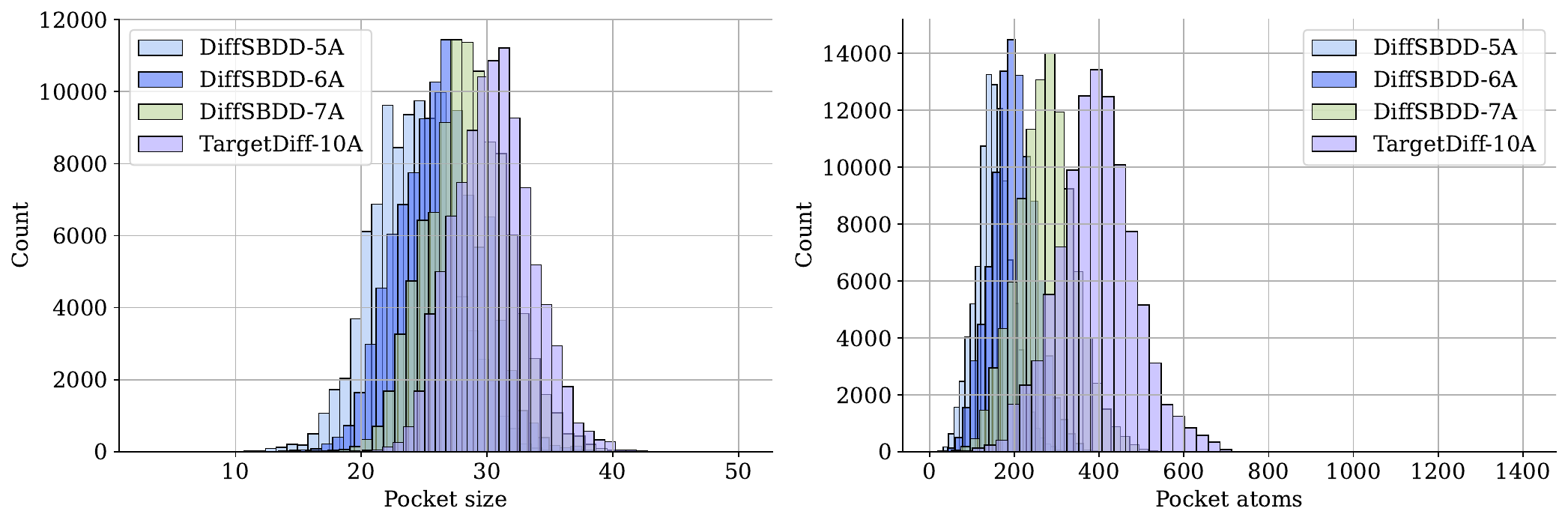}
\caption{The size distribution for protein pockets in the CrossDocked2020 dataset based on the cutoff radius to create the ligand-pocket complex. The namings DiffSBDD or Targetdiff folowed by $x$A describe the method to determine the protein pocket with cutoff $x$.
Left: size distribution. Right: distribution of pocket atoms.}
\label{fig:pocket-sizes}
\end{figure}

\begin{table}[htb!]
    \centering
        \begin{tabular}{lrrrr}
        \toprule
        Dataset & Mean & Standard Deviation & Median & Skewness \\
        \midrule
        DiffSBDD-5\angstrom & 24.16 & 3.35 & 24.20 & 0.02 \\
        DiffSBDD-6\angstrom & 26.22 & 3.33 & 26.26 & 0.08 \\
        DiffSBDD-7\angstrom & 28.47 & 3.25 & 28.40 & 0.13 \\
        TargetDiff-10\angstrom & 30.49 & 3.10 & 30.47 & 0.17 \\
        \bottomrule
        \end{tabular}
    \caption{Statistics of pocket size for different datasets.}
    \label{tab:stats-1-cd}
\end{table}

\begin{table}[htb!]
    \centering
        \begin{tabular}{lrrrr}
        \toprule
        Dataset & Mean & Standard Deviation & Median & Skewness \\
        \midrule
        DiffSBDD-5\angstrom & 152.73 & 42.04 & 151.00 & 0.26 \\
        DiffSBDD-6\angstrom & 198.62 & 51.90 & 197.00 & 0.17 \\
        DiffSBDD-7\angstrom & 274.94 & 68.32 & 274.00 & 0.19 \\
        TargetDiff-10\angstrom & 393.83 & 90.70 & 394.00 & 0.15 \\
        \bottomrule
        \end{tabular}
    \caption{Statistics of number of pocket atoms for different datasets.}
    \label{tab:stats-2-cd}
\end{table}

To create the Protein-Ligand (PL) complex, the cutoff radius determines which protein atoms should be included next to all ligand atoms, to build the PL complex. 

The work by \citet{schneuing_2023_structurebased} in DiffSBDD creates the PL-complex by computing for each atom in each residue in the protein all pairwise distances to the ligand atoms. As long as one distance from the residues' atom is below the defined cutoff to any ligand atom, the entire residue is included into the protein pocket. Hence choosing such selection through the minimum function potentially creates larger PL complexes, but ensures that all interactions between protein-atoms to the ligands are considered.

The PL complex creation in TargetDiff \cite{guan2023d} chooses a query point from each residue through the center of mass. Based on this query point the distance to all ligand atoms are computed and the residue with its atoms are included into the PL complex, if any distance between the query point to any ligand atom is below the cutoff. Note that by computing the CoM in the first place, an initial reduction has already been done which can lead so fewer interactions and hence smaller PL complexes. The authors of TargetDiff estimate the pocket size by computing the top 10 farthest pairwise distances of protein atoms. Based on that, they select the median of that as the pocket size for robustness. As the cutoff increases, we observe that the protein pocket also increases as shown in the both panels of Figure \ref{fig:pocket-sizes} for pocket size as well as the number of atoms in the protein pocket. The TargetDiff-10\angstrom~ PL-dataset is particularly large with protein pockets having mean size of 393 atoms. Having larger PL complexes might impede the optimization of the diffusion models since smaller batch sizes and backbones with less trainable parameters are only feasible to fit on a single GPU. Additionally, a smaller cutoff also enables faster training since less message passing steps are required. We believe that a cutoff of 7\angstrom~ is a good trade-off, in that it enables the diffusion model to propagate distant information but also does not fall into the risk of overfitting on a potentially smaller PL complex. The latter is particularly important in the setting of generalization when the diffusion model generates ligands on a new protein target. 

\subsection{Metrics dependency on ligand size}

\begin{figure}[htb!]
\centering
\includegraphics[width=\textwidth]{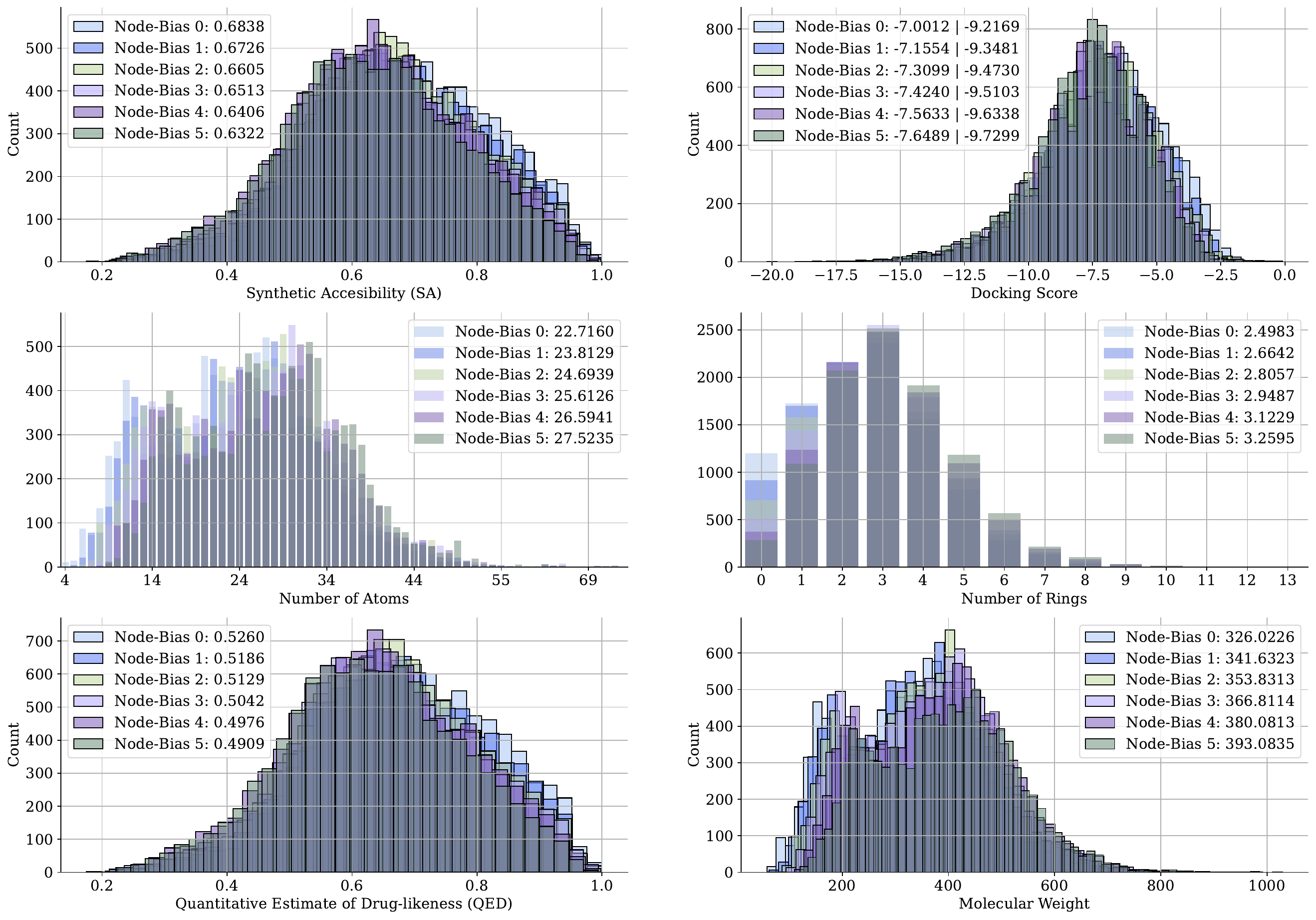}
\caption{Evaluation metrics for $10,000$ ligands where the number of atoms was first sampled from the training set prior and additionally a node bias of $n$ added. Here $n=\{0, 1, 2, 3, 4, 5\}$. The caption shows the mean value across each of the $5$ different sets per metric. For the docking scores panel, the first value describes the mean value among all targets, while the second value refers to the top-10\% mean value among all targets.}
\label{fig:evaluation-ligand-size}
\end{figure}

In this section we show how several metrics are dependent on the ligand size. To this end, we ran 6 additional sample-experiments with the on CrossDocked2020 trained PILOT-model, where ligands are generated without any guidance, i.e., unconditionally with only the protein pocket as context. Figure \ref{fig:evaluation-ligand-size} shows the results. In particular, as described in the main text, larger ligands with more atoms, shown through increasing node bias $n$ tend to have lower synthetic accesibility (SA) score as well as QED and a better (smaller) docking score.
Therefore, it is important to take the ligand size distribution into account when evaluating generative models based on docking scores, since the later is negatively correlated with ligand size.

\subsection{Ring Distribution}

\begin{figure}[htb]
\centering
\includegraphics[width=1.\textwidth]{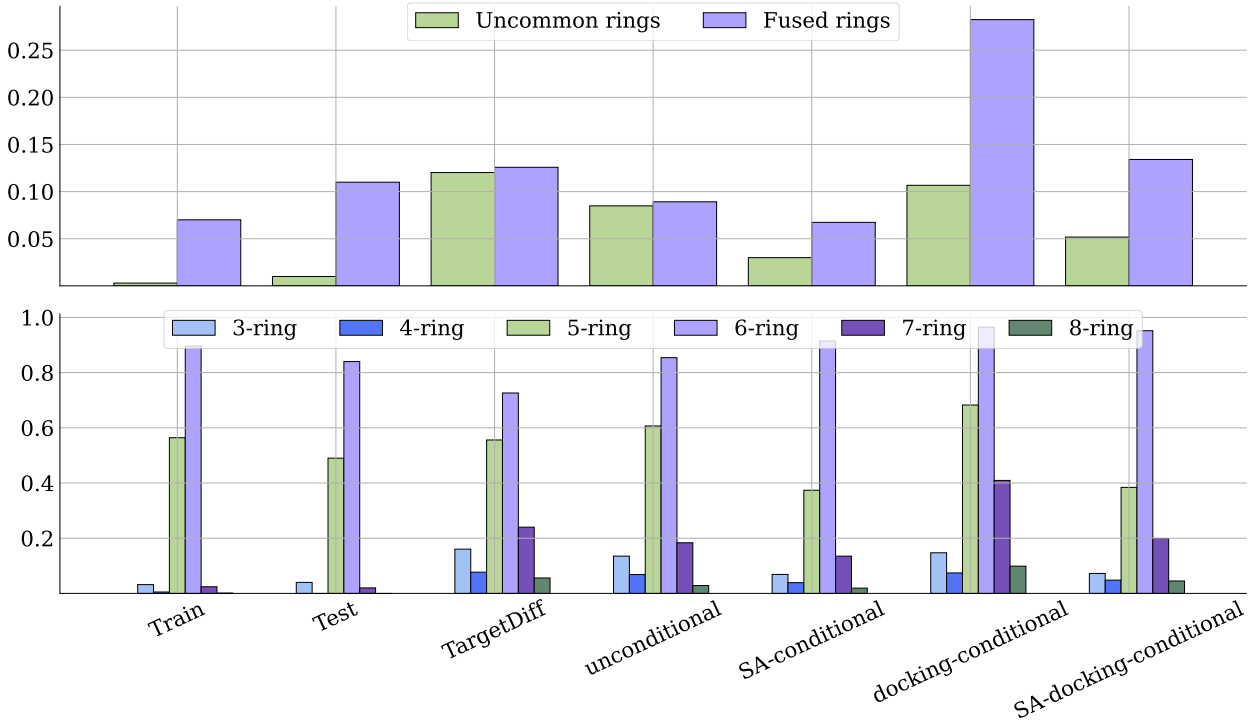}
\caption{Evaluation of the ring systems in all sampled ligands across test targets. \textbf{Left}: Histogram detailing the percentage of uncommon and fused rings for all ligands. \textbf{Right}: Histogram displaying the distribution of ring sizes from three- to eight-membered rings. Five- and six-membered rings are considered the most drug-like.}
\label{fig:eval_rings_crossdocked}
\end{figure}

To delve deeper into our analysis, we also examine the distribution of ring structures, a known challenge for 3D-based models.\citep{tamgen} The top panel in Fig. \ref{fig:eval_rings_crossdocked} illustrates the occurrence of fused and uncommon rings for all models. We observe that TargetDiff, as well as our models, tend to generate more uncommon rings compared to the train and test sets. However, both the SA- and SA-docking-conditional models effectively mitigate this issue by reducing the number of uncommon rings and aligning more closely with the distribution observed in the training and test data.

Consistent with our earlier discussion, the docking-conditional model exhibits a strong propensity for generating numerous rings, including fused and uncommon ones. As depicted in the lower panel of Fig. \ref{fig:eval_rings_crossdocked}, all models also tend to produce rings that are less common in drug-like molecules, such as three-, four-, seven-membered, or larger rings. These ring structures are often associated with poor synthetic accessibility, chemical stability, toxicity, or metabolic instability. \citep{rings_in_drugs,7membered, heterocycles}

In contrast, five- and six-membered heterocycles containing one or more heteroatoms are considered the gold standard in drug-like molecules \cite{rings_in_drugs, heterocycles, heterocycles_2}, and we observe that these are well represented in the sample space following the training distribution.

Notably, the SA-conditional model effectively regulates the formation of unfavorable ring systems, particularly three- and seven-membered rings. Conversely, the SA-docking-conditional model strikes a reasonable balance, with only a slight increase in seven-membered rings compared to the docking-conditional model, where such rings are more prevalent.

\section{Hyperparameters for Importance Sampling}
In Algorithm \ref{algorithm:importance-sampling}, we present the property-guided sampling algorithm, which we will further discuss in this section. 
To perform SA- and docking-score optimization on CrossDocked, we first optimize the SA-score in the population for several iterations. After this initial optimization, we then optimize for docking-score, using a population that has been filtered or biased based on the SA-score guidance. 
In Table \ref{table:importance-sampling-joint}, we report the start and end points, as well as the temperature parameter for the unbounded SA-score maximization and unbounded docking-score minimization. Note that the reverse diffusion trajectory sampling includes \( T = 500 \) timesteps. We set the population size to 40. This means that for a given protein target \( P \), each batch consists of 40 ligands optimized for high SA-scores and low docking scores. We continue the sampling process until 100 valid ligands are generated.

\begin{table}[htb!]
    \centering
        \begin{tabular}{lrrrr}
        \toprule
        Property & Start &  End & $N$ & Temperature $\tau$ \\
        \midrule
        SA-score & 0 & 200 & 5 & 0.1 \\
        Docking-score & 200 & 300 & 5 & 0.1 \\
        \bottomrule
        \end{tabular}
    \caption{Settings for importance sampling to optimize SA- and docking scores. Start and end columns determine when the importance sampling is performed in the iteration ranging from $1$ to $500$.}
    \label{table:importance-sampling-joint}
\end{table}

In our experiments, we tried optimizing both SA- and docking score simultaneously by either pointwise adding or multiplying the importance weights \(\{w_{k, \text{SA}}\}_{k=1}^K\) with \(\{w_{k, \text{dock}}\}_{k=1}^K\) for each intermediate ligand $k$ for varying time intervals including \((1, 200)\), \((100,300)\) and \((300,500)\). We did not see satisfying results where both criteria are optimized in the final ligands, which might be possible to achieve through different hyperparameters including temperature annealing. For this reason, we choose to optimize each property in consecutive order, where the SA-guidance shall act as an initial filter to discard synthetic infeasible (noisy) ligands between steps \((100, 250)\), while the filtered noisy ligands are then
guided to minimize docking score in the interval \((300, 400)\). Notice that we do not employ the guidance in iteration in the intervals, but every 10, such that each SA- and docking-guidance include 15 and 10 guidance steps respectively.

\section{Kinodata-3D: Analysis}
\subsection{Correlation}
\begin{figure}[htb]
\centering
\includegraphics[width=1.\textwidth]{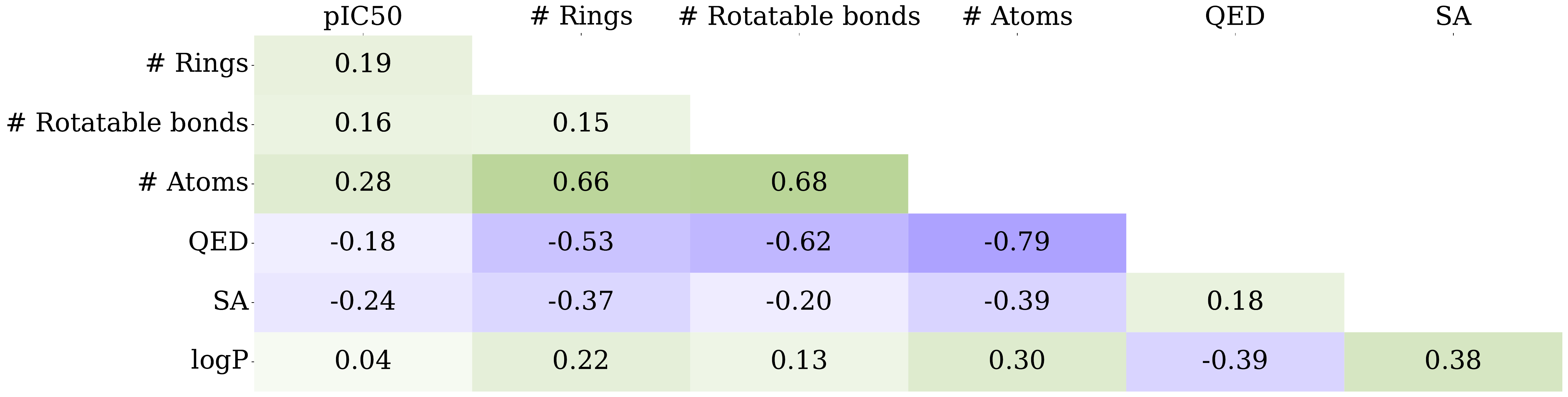}
\caption{Correlation matrix of pIC50s, number of rings, number of atoms, QEDs, and SAs on the Kinodata-3D training set.}
\label{fig:correlation-kinodata}
\end{figure}

We show the correlation matrix on the Kinodata-3D dataset in Figure \ref{fig:correlation-kinodata}. Similar to the CrossDocked dataset, we observe that metrics like the QED and SA score are negatively correlated with the number of atoms and rings. As opposed to CrossDocked, in the Kinodata-3D a negative correlation of -0.39 between logP and QED is observed while in CrossDocked the correlation amounts to 0.36. One possible explanation for this is that Kinodata-3D consists of experimental kinase-ligand assay data and therefore only considers ligands that covers a smaller chemical space where the logP covers a potentially smaller range.

\begin{figure}[htp!]
\centering
\includegraphics[width=1.\textwidth]{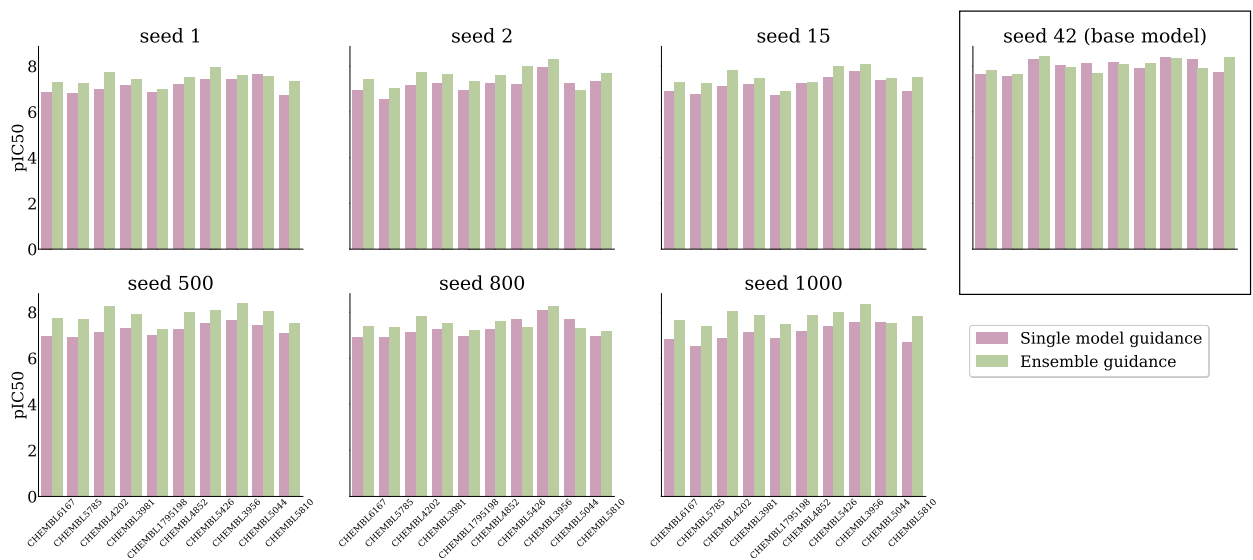}
\caption{Single model guidance is compared to ensemble guidance. A base model, here the model with seed 42, is used to sample 100 ligands per target. In the case of single model guidance (seed 42 model guides itself), a variety of models trained with different seeds evaluate the sampled ligands with respect to p\ic~values. In the case of ensemble guidance, an ensemble of models, here seed 42, 500, and seed 1000, are used for p\ic~guidance. Again, all seed models evaluate the sampled ligands with respect to p\ic. We can see that throughout targets and seeds, the ensemble guidance not only works best in terms of p\ic~but also in terms of stability and generality. The base model assigns similar p\ic~values to its samples for both, single model and ensemble guidance. Nevertheless, across seed models (involved in ensemble guidance or not) the samples taken from the single model guidance exhibit a significantly worse p\ic~prediction in contrast to ensemble guidance. Here, all seed models predict similarly high p\ic~values suggesting that ensemble guidance leads to a more stable and most importantly more general set of samples.}
\label{fig:kinodata_ensemble}
\end{figure}

\end{document}